\documentclass[11pt,a4paper]{article}
\usepackage{jheppub}
\usepackage{verbatim}
\usepackage{float}
\usepackage{subcaption}
\usepackage[utf8]{inputenc}

\newcommand{\be}{\begin{equation}}
\newcommand{\ee}{\end{equation}}
\newcommand{\bea}{\begin{eqnarray}}
\newcommand{\eea}{\end{eqnarray}}
\newcommand{\bean}{\begin{eqnarray*}}
\newcommand{\eean}{\end{eqnarray*}}

\def\beq{\begin{equation}}

\def\eeq{\end{equation}}
\def\R{\mathcal{R}}

\def\Re{\mathop{\rm Re}}
\def\Im{\mathop{\rm Im}}
\relax

\def\d{\partial}

\def\a{\alpha'}

\title{Asymptotic quasinormal modes of string-theoretical \boldmath $d$-dimensional black holes}

\author[a]{Filipe Moura,}
\author[b]{Jo\~ao Rodrigues}

\affiliation[a]{Departamento de Matem\'atica, Escola de Tecnologias e Arquitetura, \\ ISCTE - Instituto Universit\'ario de Lisboa \\ and Instituto de Telecomunica\c c\~oes, \\Av. das For\c cas Armadas, 1649-026 Lisboa, Portugal}

\affiliation[b]{
Centro de F\'\i sica da Universidade de Coimbra,\\ Rua Larga, 3004-516 Coimbra, Portugal}

\emailAdd{fmoura@lx.it.pt}
\emailAdd{jvrodrigues@pollux.fis.uc.pt}

\abstract{We compute the quasinormal frequencies of $d$-dimensional spherically symmetric black holes with leading string $\a$ corrections for tensorial gravitational perturbations in the highly damped regime. We solve perturbatively the master differential equation and we compute the monodromies of the master perturbation variable (analytically continued to the complex plane) in different contours, in order to obtain the quasinormal mode spectra. We proceed analogously for the quasinormal modes of test scalar fields. Differently than in Einstein gravity, we obtain distinct results for the two cases.
}

\arxivnumber{2105.02616}

\allowdisplaybreaks
\begin{document}
\maketitle





\section{Introduction}
\noindent

Black hole quasinormal modes (QNMs) are associated to perturbations that can be related either to the black hole metric (gravitational perturbations) or to external fields. They represent resonances in black hole scattering problems that are purely ingoing at the event horizon and purely outgoing at infinity. These boundary conditions imply that the associated frequencies must be complex. The quasinormal ringing frequencies carry unique information about parameters of the black hole in the ringdown phase resulting from a black hole collision; they can be directly measured by the gravitational wave detectors. This feature turns QNMs into preferential probes for testing theories of gravity beyond Einstein, since the ringing frequencies represent a universal part of the gravitational wave signals. These frequencies do not depend on what drives the perturbations: they are given exclusively in terms of intrinsic physical quantities of the black hole such as mass, charge or spin, and eventually (beyond Einstein gravity) some other parameters of the theory. With the advent of gravitational wave astronomy, therefore, interest in the study of black hole QNMs has raised. 
Literature concerning QNMs is vast and rich; many excellent references can be provided, such as the seminal works  \cite{Konoplya:2011qq,Konoplya:2003ii,Konoplya:2002zu,Berti:2009kk}. More recent works have extended the calculation of QNMs in several different black hole solutions \cite{Rincon:2018sgd,Panotopoulos:2017hns,Destounis:2018utr,Destounis:2020pjk,Panotopoulos:2018hua}.

Most of the times, QNMs have to be computed numerically. Nonetheless, different analytical methods have been developed in order to compute QNMs in some limiting cases. One of such cases is the asymptotic (highly damped) limit. Quasinormal frequencies in that limit have been computed in \cite{Motl:2003cd, Birmingham:2003rf, Andersson:2003fh} for $d$ dimensional asymptotically flat spherically symmetric black holes. These results have been extended to rotating black holes in $d=4$ \cite{Musiri:2003ed} and to nonasymptotically flat black holes: in $d=4$ in \cite{Cardoso:2004up} and, for $d>4$, in \cite{Musiri:2003rs, Natario:2004jd}. Numerical studies have confirmed the results of these calculations, and have also been extended to other solutions \cite{Nollert:1993zz, Cardoso:2003cj, Berti:2003zu, Berti:2003jh}.

From what we mentioned, it is relevant to extend these results to theories beyond Einstein gravity, namely theories with higher derivatives. There are several recent works computing QNMs of black holes with higher derivative corrections coming from different theories, namely \cite{Blazquez-Salcedo:2016enn, Cano:2020cao, Pierini:2021jxd, Moura:2021eln}. Concerning specifically the asymptotic limit, numerical results have been obtained in \cite{Daghigh:2006xg} for vectorial perturbations of $d$-dimensional spherically symmetric black holes in Gauss-Bonnet gravity.

In this article we will consider $d$-dimensional spherically symmetric black holes with leading string-theoretical $\a$ corrections, and analytically compute their quasinormal frequencies in the asymptotic (highly damped) limit corresponding to tensorial gravitational perturbations and test scalar fields. The article is organized as follows. In section \ref{sstp} we will review the tensorial gravitational perturbations of spherically symmetric black holes in $d$ dimensions. We then concentrate on the effective action of bosonic/heterotic superstrings with leading $\a$ corrections and on a $d$-dimensional black hole solution of its field equations. We write down the master equation and the respective potential corresponding to these perturbations. Given this information, in section \ref{qnm} we compute the quasinormal spectrum in the highly damped limit corresponding to these perturbations. We solve the master differential equation perturbatively, considering the radial variable $r$ (or more precisely the tortoise coordinate) to be complex. We consider two different (but homotopic) contours in the complex $r$ plane, associating to each of them one of the quasinormal modes' boundary conditions. By computing the monodromies of the master perturbation variable along each of these contours and equating such monodromies, we are able to obtain a condition that we can solve for the black hole quasinormal frequencies. In section \ref{testas} we proceed analogously with test scalar fields in the background of the same black hole, and we compute the respective quasinormal spectrum in the highly damped limit. In the end we discuss and compare our results in both cases.

\section{String-corrected spherically symmetric black holes and their tensorial perturbations}
\label{sstp}
\noindent

A general static spherically symmetric metric in $d$ dimensions can always be cast in the form
\be \label{schwarz}
ds^2 = -f(r)\ dt^2  + f^{-1}(r)\ dr^2 + r^2 d\Omega^2_{d-2}.
\ee

General tensors of rank at least 2 on the $(d-2)$-sphere $\mathbb{S}^{d-2}$ can be uniquely decomposed in their tensorial, vectorial and scalar components. That is the case of general perturbations $h_{\mu\nu}=\delta g_{\mu\nu}$ of a $d-$dimensional spherically symmetric metric like (\ref{schwarz}). We have then scalar, vectorial and (for $d>4$) tensorial gravitational perturbations.

Each type of perturbation is described in terms of a master variable. In Einstein gravity, each of these master variables obeys a second order differential equation (``master equation'') with a potential that depends on the kind of perturbation one considers \cite{ik03a}.

Tensorial perturbations are expressed in terms of the eigentensors $\mathcal{T}_{ij}$ of $D^2$, with $D$ being the covariant derivative on the $(d-2)$-sphere $\mathbb{S}^{d-2}$:
$$\left(D^k D_k + \ell \left( \ell + d - 3 \right) -2 \right) \mathcal{T}_{ij} \left(\theta^k \right)= 0, \, \, \ell = 2,3,4,\ldots$$
Here $i, j, k = 1,\ldots, d-2$ represent coordinates of $\mathbb{S}^{d-2}$ and $\ell$ the multipole number. $\mathcal{T}_{ij}$ also satisfy
$$D^i \mathcal{T}_{ij}=0, \, g^{ij} \mathcal{}T_{ij}=0.$$

Specifically, tensorial gravitational perturbations of the metric (\ref{schwarz}) are given in terms of a function $H_T (r,t)$ by
\bea
h_{ir}, \, h_{it}, \, h_{rr}, \, h_{rt}, \, h_{tt}=0, \nonumber \\
h_{ij}=2 r^2 H_T (r,t) \mathcal{T}_{ij}. \label{tpert}
\eea
Assuming an oscillatory time dependence, from the perturbation function $H_T (r,t)$ we can define a master variable $\psi(r)$ through
\be
H_T (r,t)= \frac{\mbox{e}^{i \omega t}}{\kappa(r)} \psi(r), \label{mv}
\ee
with $\kappa(r)=r^{\frac{d-2}{2}}.$
In terms of the tortoise coordinate $x$ for the metric (\ref{schwarz}) defined by
\be
dx = \frac{dr}{f(r)}, \label{tort}
\ee
the master variable $\psi(r)$ satisfies a second order differential equation with a potential, the ``master equation'', given by
\be
\frac{d^2 \psi}{d\, x^2} + \omega^2 \psi = V \left[ f(r) \right] \psi \label{potential0}
\ee
with the potential $V \left[ f(r) \right]$ in this case being given by the minimal potential
\be
V_{\textsf{min}} [f(r)] = f(r) \left( \frac{\ell \left( \ell + d - 3 \right)}{r^2} + \frac{\left( d - 2 \right) \left( d - 4 \right) f(r)}{4r^2} + \frac{\left( d - 2 \right) f'(r)}{2r} \right). \label{v0}
\ee

In the presence of higher order corrections in the lagrangian, one can still have spherically symmetric black holes of the form (\ref{schwarz}), but the master equation obeyed by each perturbation variable is expected to change. Concretely, we will consider the following $d$--dimensional effective action with string $\a$ corrections:
\be \label{eef} \frac{1}{16 \pi G} \int \sqrt{-g} \left( \R -
\frac{4}{d-2} \left( \d^\mu \phi \right) \d_\mu \phi +
\mbox{e}^{\frac{4}{d-2} \phi} \frac{\lambda}{2}\
\R^{\mu\nu\rho\sigma} \R_{\mu\nu\rho\sigma} \right) \mbox{d}^dx .
\ee
This is the effective action of bosonic and heterotic string theories, to first order in the inverse string tension $\a$, with $\lambda = \frac{\a}{2}, \frac{\a}{4}$, respectively. \footnote{Type II superstring theories do not have $\a$ corrections to this order.} In both cases, since we are only interested in purely gravitational corrections, we can consistently set all other bosonic and fermionic fields present in the string spectrum to zero except for the dilaton field $\phi$.

Tensorial metric perturbations in the presence of these leading string $\a$ corrections for a spherically symmetric metric of the form (\ref{schwarz}) are also defined through (\ref{tpert}). The master variable $\psi$ is also defined through (\ref{mv}), with
$$\kappa(r)=\frac{1}{\sqrt{f}} \exp\left(\int
\frac{\frac{f'}{f} +\frac{d-2}{r} + \frac{4}{r^3} (d-4) \lambda
(1-f) - \frac{4}{r^2} \lambda f' - \frac{2}{r f} \lambda
f'^2}{2-\frac{4}{r}\lambda f'} dr\right).$$
In \cite{Moura:2006pz,Moura:2012fq} it has been shown that, perturbing the field equations resulting from this action, for tensorial perturbations of the metric (\ref{schwarz}) one also obtains a second order master equation like (\ref{potential0}). The corresponding potential, as expected, is an $\a$-corrected version of the minimal potential in (\ref{v0}) given by
\bea
V_{\textsf{T}} [f(r)] &=& \lambda\ \frac{f(r)}{r^2} \left[ \left( \frac{2 \ell \left( \ell + d - 3 \right)}{r} + \frac{\left( d - 4 \right) \left( d - 5 \right) f(r)}{r} + \left( d - 4 \right) f'(r) \right) \left( 2 \frac{1 - f(r)}{r} + f'(r) \right) \right. \nonumber \\
&+& \left. \Big( 4 (d-3) - (5d-16) f(r) \Big) \frac{f'(r)}{r} - 4 \left( f'(r) \right)^2 + \left( d-4 \right) f(r) f''(r) \right] + V_{\textsf{min}} [f(r)]. \label{vt}
\eea

Spherically symmetric $d-$dimensional black hole solutions with leading $\a$ corrections have been obtained in \cite{cmp89,Moura:2009it}. Specifically concerning the action (\ref{eef}), a solution of the respective field equations is of the form (\ref{schwarz}), with

\bea \label{fr2}
f(r) &=& f_0(r) \left(1+ \frac{\lambda}{R_H^2} \delta f(r) \right), \\
f_0(r) &=& 1 - \frac{R_H^{d-3}}{r^{d-3}}, \label{fr0}\\
\delta f(r) &=& - \frac{(d-3)(d-4)}{2}\ \frac{R^{d-3}_H}{r^{d-3}}\ \frac{1 - \frac{R_H^{d-1}}{r^{d-1}}}{1 - \frac{R^{d-3}_H}{r^{d-3}}}.
\eea
The only horizon of this metric occurs at the same radius $r=R_H$ of the Tangherlini solution, which is the metric with $f(r)=f_0(r)$ obtained in the Einstein limit $\lambda=0$.

This black hole solution has been obtained by Callan, Myers and Perry in \cite{cmp89}, where some of its properties have been studied. For our purposes, it is enough to quote here the explicit expression for its temperature, given by
\be
T_{\mathcal{H}} = \frac{f'(R_H)}{4 \pi}= \frac{f_0'(R_H)}{4 \pi} \left(1+ \frac{\lambda}{R_H^2} \delta f(R_H) \right)= \frac{d-3}{4 \pi R_H} \left( 1 - \frac{\left( d-1 \right) \left( d-4 \right)}{2}\ \frac{\lambda}{R_H^2} \right). \label{temp}
\ee
Throughout this article we will use for the perturbative expansion the small dimensionless parameter
\be
\lambda' = \frac{\lambda}{R_H^2}= \lambda \frac{16 \pi^2}{(d-3)^2} T_\mathcal{H}^2. \label{lprime}
\ee

The stability of this solution under tensorial gravitational perturbations has been studied in \cite{Moura:2006pz}, and the spectra of quasinormal modes corresponding to such perturbations and also of test scalar fields in the eikonal limit has been obtained in \cite{Moura:2021eln}. In this article we will compute the same spectra of quasinormal modes for both cases in the highly damped regime.

\section{Asymptotic quasinormal modes of tensorial gravitational perturbations to the Callan-Myers-Perry black hole}
\label{qnm}
\noindent

\subsection{Quasinormal modes, boundary conditions and the monodromy method}
\label{bcqm}

Quasinormal modes of tensorial gravitational perturbations are solutions to the corresponding master equation (\ref{potential0}) subject to the boundary conditions
\bea
\psi \propto e^{-i\omega x} \hspace{3pt} , \hspace{3pt} r \to +\infty;
\label{bcr1} \\
\psi \propto e^{i\omega x} \hspace{3pt} , \hspace{3pt} r \to R_H.
\label{bcr2}
\eea
In terms of the tortoise coordinate $x$, these boundary conditions are written as
\bea
\psi \propto e^{-i\omega x} \hspace{3pt} , \hspace{3pt} x \to +\infty;
\label{bcx1} \\
\psi \propto e^{i\omega x} \hspace{3pt} , \hspace{3pt} x \to -\infty.
\label{bcx2}
\eea

It should be clear that any method devised to compute quasinormal frequencies will have to make use of these defining boundary conditions. Because quasinormal frequencies are complex and $r$ and $x$ are real, gathering information of these boundary conditions amounts to distinguish between an exponentially vanishing and an exponentially growing term. Clearly, a numerical approach will face problems with such task. Moreover, any kind of analytical approximate approach will also fail to some extent. Indeed, many lower order terms of the approximation are needed in order to make sense of the exponentially decreasing term, otherwise this term might be much smaller than the approximation error and consequently needs to be disregarded.

An elegant solution to half of this issue emerges from the moment we allow $r$ (and $x$) to take complex values and consequently assume an analytic continuation of functions of $r$ to the complex plane. Indeed, if one takes the contour $\Im\left(\omega x\right)=0$ in the complex $r$ plane, then
$|e^{\pm i\omega x}| = 1,$ the asymptotic behavior of $e^{\pm i\omega x}$ is always oscillatory and there will be no problems with exponentially growing versus exponentially vanishing terms. Thus if one considers the Stokes lines $\Im\left(\omega x\right)=0$, imposing the boundary condition (\ref{bcx1}) in the complex $r$ plane no longer poses a challenge to an approximate analytical method.

Near the event horizon, we can distinguish the two exponential terms by computing the respective monodromies around it. As we will see in the next section, these monodromies are non trivial because $x$ has a branch point in the event horizon. As it turns out, the boundary condition (\ref{bcx2}) can be set as a monodromy condition.

The monodromy method we will be using was introduced in \cite{Motl:2003cd}. In order to apply it one needs two contours, and the solutions to the master equation in the respective regions, to impose the appropriate boundary conditions. The general idea goes as follows:

\begin{itemize}

    \item We pick two closed homotopic contours on the complex $r$-plane. Both these contours enclose only the physical horizon: none of them encloses the origin of the complex $r$-plane nor any fictitious horizon.
    \item One of these contours, which we designate as the big contour, seeks to encode information of the boundary condition (\ref{bcx1}) on the monodromy of $\psi$ associated with a full loop around it.
    \item The other contour, which we designate as the small contour, seeks to encode information of the boundary condition (\ref{bcx2}) on the monodromy of $\psi$ associated with a full loop around it.
    \item As both contours are homotopic, the monodromy theorem asserts that the respective monodromies must be the same. Thus, equating them hopefully yields a restriction on the values of the quasinormal frequencies $\omega$, from the complex plane to an infinite but countable subset.
\end{itemize}

We restrict our analysis in this article to the highly damped regime of quasinormal modes defined by the condition
\be
\Im\left(\omega\right) \gg \Re\left(\omega\right). \label{asy}
\ee
This condition is equivalent to $\omega$ being approximately imaginary. The definition of a Stokes line comes thus as
\begin{equation}
    \Im\left(\omega x\right) = 0 \Rightarrow \Re\left(x\right) =0.
\end{equation}

\subsection{Choice of coordinates}
\label{coord}

\subsubsection{Behavior close to the origin}
\noindent

The tortoise coordinate (\ref{tort}) for the metric (\ref{fr2}) we are working with is given, up to an integration constant $C_X$, in terms of the Gauss hypergeometric function $_{2}F_{1}$ by
\bea
x&=&_{2}F_{1}\left(1,-\frac{1}{d-3};1-\frac{1}{d-3};\left(\frac{R_H}{r}\right)^{d-3}\right) r - \left[\left(\frac{R_H}{r}\right)^{d-3} \frac{_{2}F_{1}\left(2,\frac{d-4}{d-3};1+\frac{d-4}{d-3};\left(\frac{R_H}{r}\right)^{d-3}\right)}{d-4} \right.\nonumber \\
&-& \left. \left(\frac{R_H}{r}\right)^{2d-4} \frac{_{2}F_{1}\left(2,\frac{2d-5}{d-3};1+\frac{2d-5}{d-3};\left(\frac{R_H}{r}\right)^{d-3}\right)}{2d-5}\right] \frac{(d-3)(d-4)}{2} \frac{\lambda}{R_H^2} r +C_X. \label{xint}
\eea

Close to the origin, this coordinate $x$ can be approximated as
\bea
x &\sim& - \frac{1}{d-2}\frac{r^{d-2}}{R_H^{d-3}}+\frac{(d-3)(d-4)}{2} \frac{\lambda}{r} \nonumber \\
&+& \mathrm{e}^{\frac{\pi i}{d-3}} \frac{\pi}{(d-3) \sin\left(\frac{\pi}{d-3}\right)} R_H + \frac{d-4}{d-3}\frac{\pi}{\sin\left(\frac{\pi}{d-3}\right)}  \frac{\lambda}{2 R_H} \left(\mathrm{e}^{\frac{\pi i}{d-3}} -(d-2) \mathrm{e}^{\frac{-\pi i}{d-3}} \right)+ C_X . \label{x0}
\eea
We see that, associated to the $\a$ correction, there is a singularity in the coordinate $x$ at $r=0$.
Because of such singularity, close to the origin the Stokes lines $\Re(x)=0$ are very difficult to handle. Since the analysis of these lines is crucial for our calculation, we must find an alternative coordinate in order to avoid that singular behavior close to the origin.

We have chosen to rather take the tortoise coordinate $z$ corresponding to the Tangherlini solution: since there are no $\a$ corrections associated to it, we thought that was the most sensible choice, considering the perturbative analysis we will make. Such coordinate is given simply by
\begin{equation}
dz = \frac{dr}{f_0(r)} \label{tort2}
\end{equation}
with $f_0(r)$ given by (\ref{fr0}). After integration one gets simply the $\lambda=0$ part of (\ref{xint}):
\be
z=_{2}F_{1}\left(1,-\frac{1}{d-3};1-\frac{1}{d-3};\left(\frac{R_H}{r}\right)^{d-3}\right) r +C_Z \label{zint}
\ee
for some integration constant $C_Z$.

Both coordinates $x, z$ are defined up to the constants $C_X, C_Z$. We may choose $C_Z$ in (\ref{zint}) in such a way that, close to the origin, $z$ can be approximated simply as
\begin{equation}
z \sim - \frac{1}{d-2}\frac{r^{d-2}}{R_H^{d-3}}. \label{z0}
\end{equation}
With this choice of $C_Z$, $z$ in (\ref{zint}) can also be written as \cite{Motl:2003cd}
\be
z = r + \frac{1}{2}\sum_{n=0}^{d-4}\frac{\exp\left(\frac{2 \pi i n}{d-3}\right)}{d-3}\log\left(1-\frac{r}{R_H} \exp\left(-\frac{2 \pi i n}{d-3}\right)\right).
\ee
We see that $z(r)$ is a multivalued function (just like $x(r)$); indeed, from (\ref{tort2}) we see that each zero of $f_0(r)$ is a branch point. There are $d-3$ zeros of $f_0(r)$:
\be
f_0(r)=0 \Leftrightarrow R_n= R_H \exp\left(\frac{2 \pi i n}{d-3}\right), \, n= 0, 1, \ldots, d-4. \label{hors}
\ee
Only the solution $R_0=R_H$ corresponds to a physical horizon; the other $d-4$ solutions are ``fictitious'' horizons.

\subsubsection{Behavior at infinity}
\noindent

At spatial infinity one can easily notice that $f(r)\sim f_0(r) \sim 1$, and therefore from (\ref{xint}), (\ref{zint}) we get, in this limit, $x \sim r + C_X$, $z \sim r + C_Z$ and
\begin{equation}
z \sim x +C_Z - C_X,
\end{equation}
i.e. in spatial infinity the coordinates $r$, $x$ and $z$ are the same up to constants. The corresponding boundary condition in this limit, equivalent to (\ref{bcx1}), is then written as
\begin{equation}
\psi \propto e^{-i\omega z} \hspace{3pt} , \hspace{3pt} z \to +\infty.
\label{bcz1}
\end{equation}
In this limit we then have
\begin{equation}
\frac{dz}{dx} \sim 1, \, \frac{d^2 z}{dx^2} \sim 0;
\end{equation}
therefore, in this region the master equation describing the perturbations can be written with respect to the variable $z$ exactly in the same way as it is written with respect to $x$, i.e. (\ref{potential0}). Also from $f(r)\sim f_0(r) \sim 1$ we see that
$V_{\textsf{min}} [f(r)]$ given by (\ref{v0}), and more generally $V_{\textsf{T}} [f(r)]$ given by (\ref{vt}), vanish. This way, (\ref{potential0}) is written in this region simply as $$\frac{d^2 \psi}{dz^2} +\omega^2 \psi=0,$$ which is compatible with the required boundary condition (\ref{bcz1}).

\subsubsection{The topology of Stokes lines}
\label{topst}
\noindent

\begin{figure}[h]
\begin{subfigure}{0.32\textwidth}
\includegraphics[width=5.2cm, height=5.2cm]{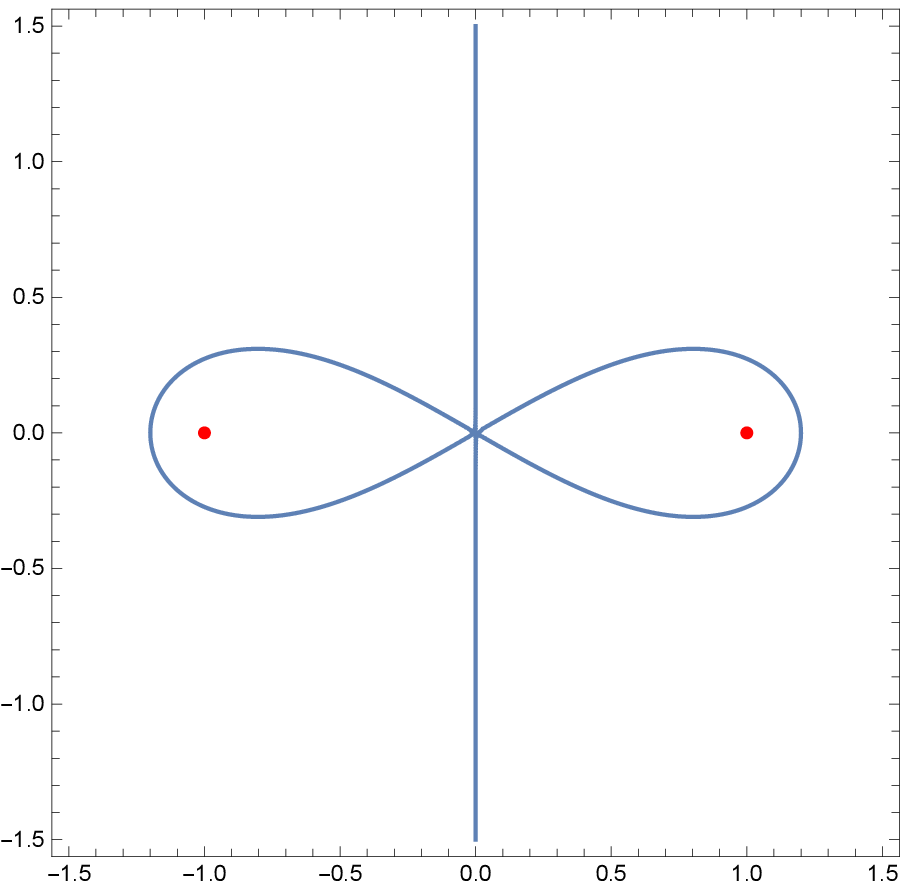}
\caption{$d=5$}
\end{subfigure}
\begin{subfigure}{0.32\textwidth}
\includegraphics[width=5.2cm, height=5.2cm]{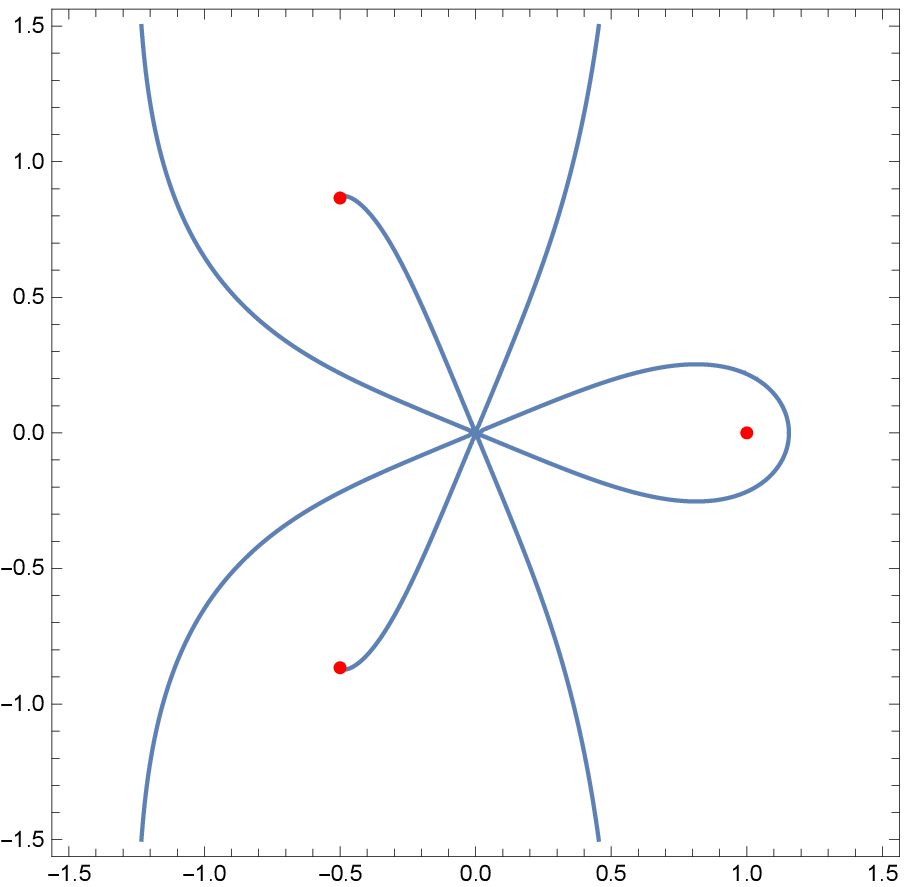}
\caption{$d=6$}
\end{subfigure}
\begin{subfigure}{0.32\textwidth}
\includegraphics[width=5.2cm, height=5.2cm]{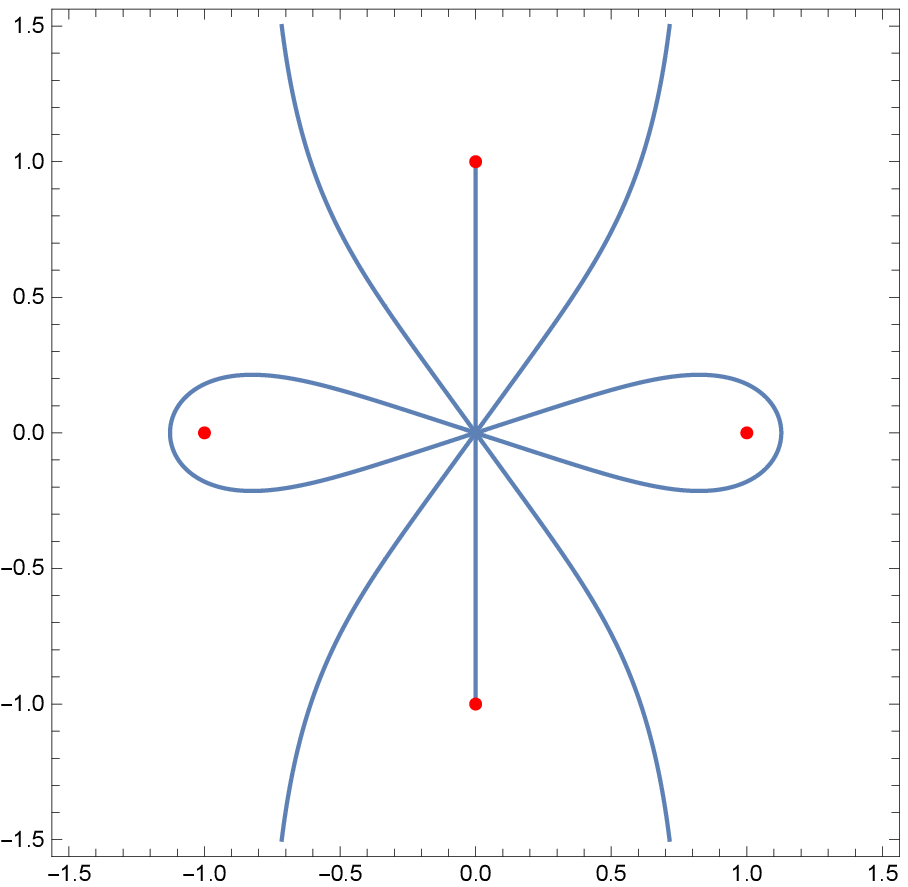}
\caption{$d=7$}
\end{subfigure}
\begin{subfigure}{0.32\textwidth}
\includegraphics[width=5.2cm, height=5.2cm]{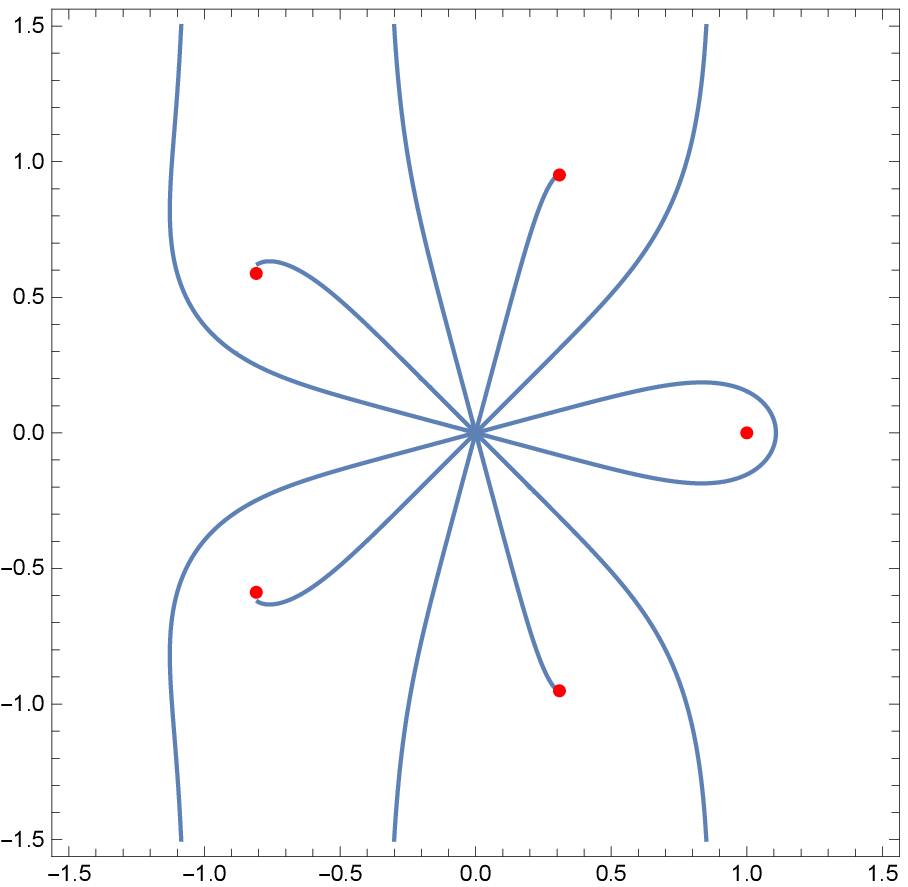}
\caption{$d=8$}
\end{subfigure}
\begin{subfigure}{0.32\textwidth}
\includegraphics[width=5.2cm, height=5.2cm]{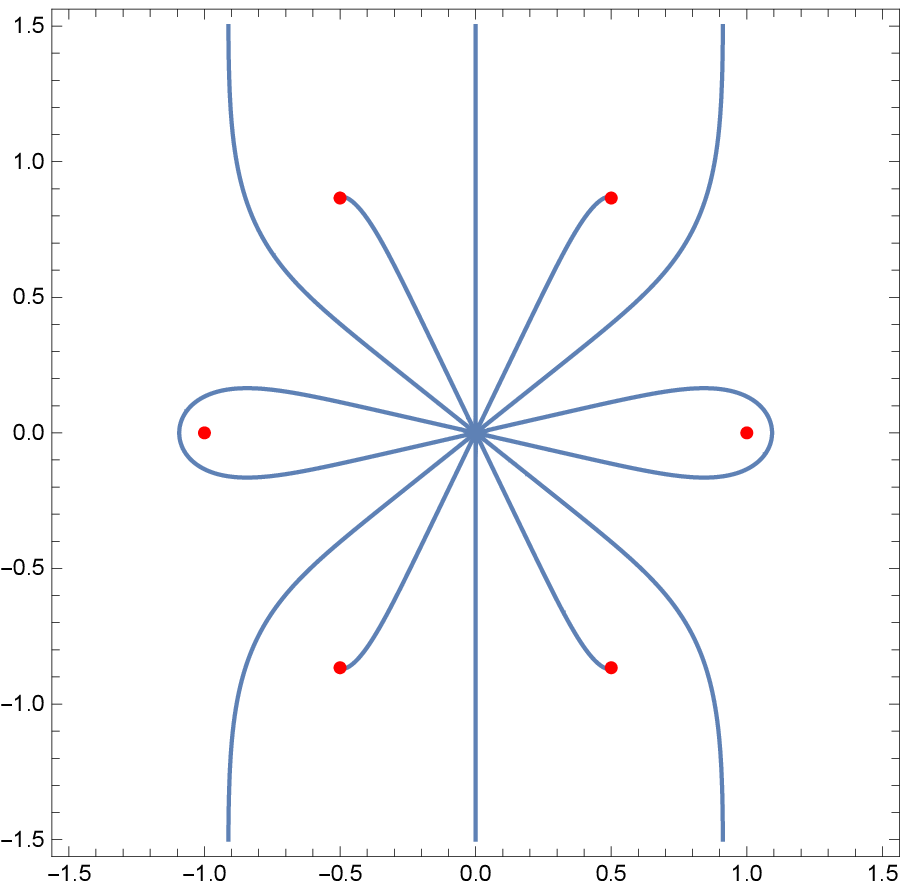}
\caption{$d=9$}
\end{subfigure}
\begin{subfigure}{0.32\textwidth}
\includegraphics[width=5.2cm, height=5.2cm]{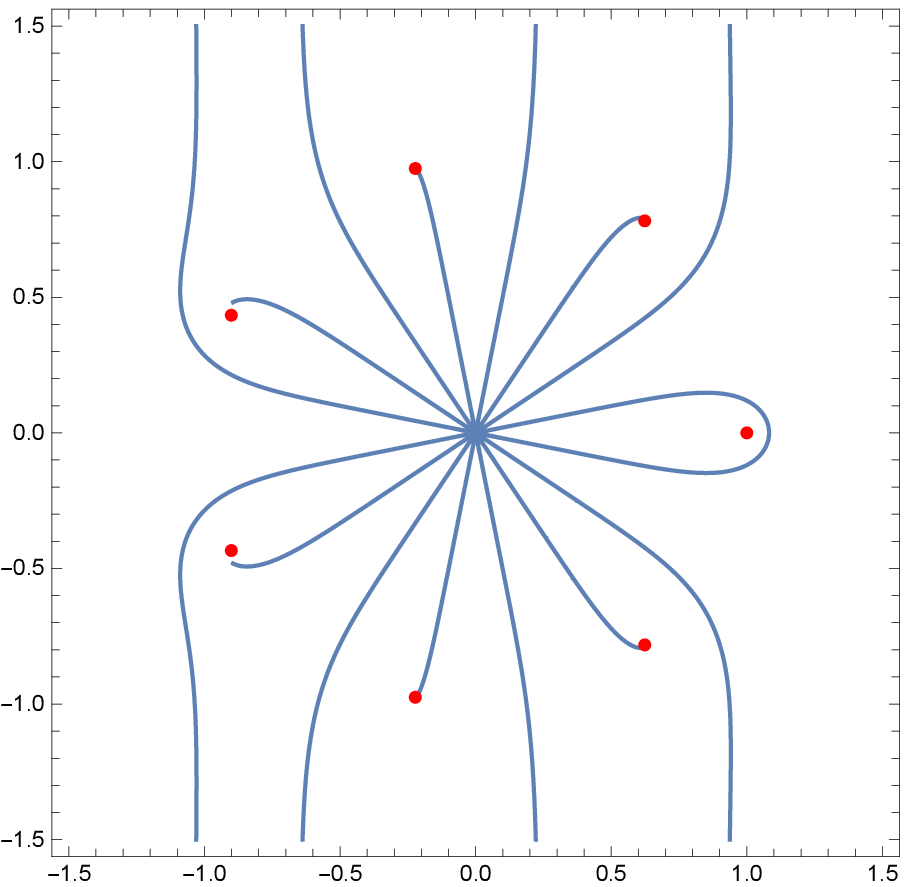}
\caption{$d=10$}
\end{subfigure}
\caption{Numerical plot of the Stokes lines topology for different dimensions. The horizontal axis stands for $\Re(r/R_H)$ and the vertical axis stands for $\Im(r/R_H)$. We denoted the positions of the physical horizon $R_H$ and of the fictitious horizons by red dots.}
\label{fig1}
\end{figure}

Since we are using the standard tortoise coordinate $z$ of the non corrected $d$-dimensional Tangherlini black hole spacetime, we will have the same structure of Stokes lines. As we saw in (\ref{z0}), close to the origin of the complex $r$-plane, the leading term of $z$ is given by $z \sim -\frac{1}{R_H^{d-3}} \frac{r^{d-2}}{d-2}.$ If we parameterize $r$ as $r = \rho e^{i\theta}$ for $\rho \in \mathbb{R}^+_0$ and $\theta \in [0,2\pi[$, we know that

\begin{equation}
    \Re \left(r^{d-2}\right) = \rho ^{d-2}\cos\left((d-2)\theta\right).
\end{equation}
Equating the expression above to zero yields
\begin{equation}
    \theta =\frac{\pi}{d-2} \left(n + \frac{1}{2}\right)
    \label{75}
\end{equation}
for $1 \leq n \leq 2d-4$. This means that in general we will have $2d-4$ Stokes lines emerging from the origin of the complex $r$-plane, all equally distributed and separated by an angle of $\frac{\pi}{d-2}$. As argued in \cite{Natario:2004jd} we know that two of such lines are bounded, forming angles of $\pm \frac{\pi}{2(d-2)}$ with the real axis at the origin, and forming a loop around the real physical horizon $r=R_H$. The next two adjacent Stokes lines are unbounded, going towards complex infinity and forming angles of $\pm \frac{3 \pi}{2(d-2)}$ with the real axis at the origin. Between these two unbounded Stokes lines there will be no ``fictitious'' horizons, as these are all located in a circumference of radius $R_H$ and the ones that are adjacent to the real horizon form angles of $\pm \frac{2 \pi}{d-3}$ with the real axis at the origin. Only for $d \geq 8$ there will be fictitious horizons with $\Re(r)>0$, but as we can see in figure \ref{fig1} these horizons lie outside the domain contained between the two unbounded Stokes lines we mentioned.

\subsection{The master equation and its perturbative solution}
\label{pert}
\noindent

We want to recast the differential equation (\ref{potential0}) with respect to the variable $z$. We may write the second order derivative in (\ref{potential0}) as
\begin{equation}
    \frac{d^2\psi}{dx^2} = \frac{d^2 \psi}{dz^2}\left(\frac{dz}{dx}\right)^2 + \frac{d\psi}{dz}\frac{d^2z}{dx^2}.
\end{equation}
From (\ref{tort}), (\ref{fr2}) and (\ref{tort2}) we get
\begin{equation}
\frac{dz}{dx} = 1+ \frac{\lambda}{R_H^2} \delta f(r); \label{dzdx}
\end{equation}
using the definition (\ref{tort}) and differentiating again (\ref{dzdx}), we obtain
\begin{equation}
    \frac{d^2z}{dx^2} = f \frac{d}{dr}\left(\frac{dz}{dx} \right) = \frac{\lambda}{R_H^2} f(r) \left(\delta f(r)\right)' \label{d2zdx2}.
\end{equation}
This way, (\ref{potential0}) is written in terms of $z$ as
\begin{equation}
\left(\frac{dz}{dx}\right)^2 \frac{d^2 \psi}{dz^2} + \frac{d^2z}{dx^2} \frac{d\psi}{dz} +\left( \omega^2 -V\right)\psi=0,
  \label{potentialz}
\end{equation}
with $\frac{dz}{dx}$ and $\frac{d^2z}{dx^2}$ given as functions of $r$ by (\ref{dzdx}) and (\ref{d2zdx2}), respectively.

Replacing (\ref{dzdx}) and (\ref{d2zdx2}), we rewrite (\ref{potentialz}) as
\be
\left[1+ \lambda' \delta f(r) \right]^2 \frac{d^2 \psi}{dz^2} + \lambda' f(r) \left[\delta f(r) \right]' \frac{d\psi}{dz} +\left( \omega^2 -V\right)\psi=0,
  \label{potentialw}
\ee

In order to solve the differential equation (\ref{potentialw}) in different regions of the complex $r$-plane, we apply standard perturbation theory.
This way we expand, to first order in $\lambda'$, the perturbation function $\psi$ and the potential $V$:
\bea
\psi &=& \psi_0 + \lambda' \psi_1, \label{psidef}\\
V(r) &=& V_0(r) + \lambda' V_1(r). \label{v1def}
\eea
$V(r)$ is the full potential $V_{\textsf{T}} [f(r)]$ given by (\ref{vt}).
The $\lambda'=0$ part is given by $V_0(r)=V_{\textsf{min}} [f_0(r)]$, with $V_{\textsf{min}}$ given by (\ref{v0}): it is the classical (uncorrected) potential evaluated with the uncorrected metric function (\ref{fr0}). All the $\lambda'$ corrections appear in $V_1(r)$: those that are implicit in $V_{\textsf{min}} [f(r)]$, from evaluating $V_{\textsf{min}}$ with a $\lambda'$-corrected function, and those that are explicit in (\ref{vt}).

Replacing the above expansions in (\ref{potentialw}) and expanding again in $\lambda'$, by separately considering the terms of order zero and first order in $\lambda'$ we obtain two separate differential equations, a homogeneous and a nonhomogeneous one:

\bea
\frac{d^2\psi_0}{dz^2} + \left(\omega^2 -V_0\right)\psi_0 &=& 0, \label{potentialz0} \\
\frac{d^2\psi_1}{dz^2} + \left(\omega^2 -V_0\right)\psi_1 &=& \xi, \label{potentialz1}
\eea
with the function $\xi$ given by
\bea
\xi &=& \xi_1+\xi_2+\xi_3, \label{xi} \\
\xi_1&=& -2 \delta f(r) \left(\frac{d^2 \psi_0}{dz^2}\right), \label{xi1} \\
\xi_2&=& - f(r) \left[\delta f(r) \right]'  \left(\frac{d\psi_0}{dz}\right), \label{xi2} \\
\xi_3&=&  V_1(r) \psi_0. \label{xi3}
\eea
The boundary conditions at spatial infinity of the differential equations (\ref{potentialz0}) and (\ref{potentialz1}) will be a simple extension of the boundary condition (\ref{bcz1}): $\psi_0(z) \sim e^{-i\omega z}, \, \psi_1(z) \sim e^{-i\omega z}, \hspace{3pt} z \to +\infty.$

\subsection{The big contour}
\label{bigc}
\noindent

In order to build the big contour, we use the properties we found when studying the topology of the Stokes lines. More precisely, for every dimension $d$, there will be two Stokes lines, emerging from the origin of the complex $r$-plane, encircling the event horizon $R_H$. Furthermore, these lines are followed, counterclockwise and clockwise, by two unbounded Stokes lines, asymptotically parallel to the imaginary axis. The big contour will follow these unbounded Stokes lines, reaching the condition $|r| \to +\infty$ twice. There, the contour abandons the Stokes lines and follows a large arc shaped path enclosing it.

\begin{figure}[h]
\centering
\includegraphics[width=0.5\textwidth]{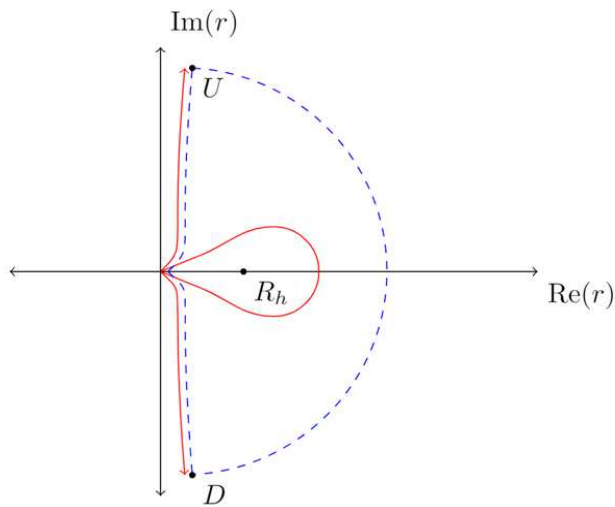}
\caption{Schematic depiction of the big contour, as the blue dashed line. The Stokes lines are depicted as red curves. Naturally, not all Stokes lines are depicted. Furthermore, we marked by $D$ and $U$ the regions where the boundary condition (\ref{bcz1}) may be imposed.}
\label{fig2}
\end{figure}

Overall, the big contour is well represented as depicted in figure \ref{fig2}. Looking at this figure, we notice the proportions may not be right. However, the topology of the big contour is well represented by the blue dashed line for every dimension $d \geq 5$.

The boundary condition (\ref{bcz1}) is to be imposed in the regions marked by $D$ or $U$. Here, we choose the region $D$ to impose it. Furthermore, we choose to follow the contour in the clockwise direction.

\begin{figure}[H]
\centering
\includegraphics[width=0.4\textwidth]{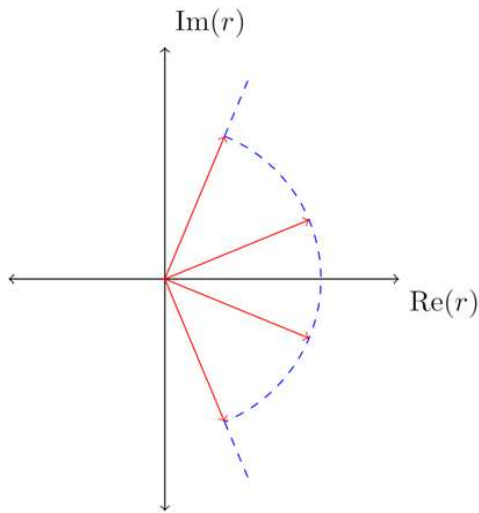}
\caption{Schematic representation of the small arc-shaped portion of the big contour, in the surroundings of $r = 0$, as the blue dashed line. The Stokes lines are represented by red curves. Naturally, not all Stokes lines are depicted.}
\label{fig3}
\end{figure}

In the surroundings of $r=0$, the big contour is represented as depicted in figure \ref{fig3}. From equation (\ref{75}) we conclude that the (small) arc shaped portion of the big contour in this region, depicted in this figure, sweeps an angle of $\frac{3\pi}{d-2}$.

\subsection{Solution in a neighborhood of the origin and asymptotic behavior}
\noindent

We will now compute the solutions to the differential equations (\ref{potentialz0}) and (\ref{potentialz1}) in a neighborhood of the origin of the complex $r-$plane. We then proceed to study the asymptotic behavior of these solutions in the portion of the big contour that follows the Stokes lines extending to complex infinity.

We start by solving (\ref{potentialz0}), the differential equation for $\psi_0$. This is precisely the same equation describing the same quasinormal modes in Einstein gravity. The resolution of this equation has been studied in detail in \cite{Motl:2003cd,Natario:2004jd}; we will review the main steps.

In an arbitrarily small neighborhood of the origin, the leading term of the potential $V_0(r)$ is given by
\be
V_0(r) \sim -\left(\frac{d-2}{2}\right)^2 \frac{R_H^{2d-6}}{r^{2d-4}}.
\ee
Replacing $z$ given by (\ref{z0}) in $V_0(r)$, we obtain for the leading term $V_0(z) \sim -\frac{1}{4z^2}$. We can then write the differential equation (\ref{potentialz0}), in this neighborhood, as
\begin{equation}
    \frac{d^2\psi_0}{dz^2} + \left(\omega^2 - \frac{j^2-1}{4z^2}\right)\psi_0 = 0 \label{psi0e}
\end{equation}
for $j =0$. Following the procedure of \cite{Motl:2003cd,Natario:2004jd} we will consider the general solution, for arbitrary $j$, of the above differential equation, and at the end take the limit $j \rightarrow 0$. Such solution is given by
\begin{equation}
    \psi_0(z) = A_+ \sqrt{2\pi} \sqrt{\omega z} J_{\frac{j}{2}}(\omega z) + A_- \sqrt{2\pi} \sqrt{\omega z} J_{-\frac{j}{2}}(\omega z), \label{psi0}
\end{equation}
where $J_{\pm\frac{j}{2}}(\omega z)$ are Bessel functions of the first kind and $A_+,A_-$ arbitrary constants.

Bessel functions verify the asymptotic condition (for $|\omega z| \gg 1$)
\bea
&&\sqrt{2\pi}\sqrt{\omega z}J_{\pm\frac{j}{2}}(\omega z) \sim 2\cos\left(\omega z -\alpha_{\pm}\right), \nonumber \\
&&\alpha_{\pm} = \frac{\pi}{4}\left(1 \pm j\right). \label{alfa}
\eea
The condition $|\omega z| \gg 1$ holds in the portion of the big contour following the Stokes lines everywhere except near the origin. Therefore, in this portion of the contour extending to complex infinity $\psi_0$ will be a linear combination of plane waves:
\bea
\psi_0 (z) & \sim & 2 A_+ \cos \left( \omega z - \alpha_+ \right) + 2 A_- \cos \left( \omega z - \alpha_- \right) \nonumber \\
& = & \left( A_+ e^{-i\alpha_+} + A_- e^{-i\alpha_-}\right) e^{i \omega z} + \left( A_+ e^{i\alpha_+} + A_- e^{i\alpha_-}\right) e^{-i \omega z}. \label{psias0}
\eea
Imposing the boundary condition (\ref{bcz1}) to $\psi_0$, we get a linear system of equations for the coefficients $A_+, A_-$ (with $c \in \mathbb{C}$):
\begin{equation}
    \begin{cases}A_+e^{i\alpha_+} + A_-e^{i\alpha_-} = c, \\
    A_+e^{-i\alpha_+} + A_-e^{-i\alpha_-} = 0,
    \end{cases}
    \label{a0}
\end{equation}
which we can solve:
\begin{equation}
    \begin{cases}
    A_- = \frac{c e^{-i\alpha_+}}{2i\sin(\alpha_- - \alpha_+)}, \\
    A_+ = -\frac{c e^{-i\alpha_-}}{2i\sin(\alpha_- - \alpha_+)}. \label{a0s}
    \end{cases}
\end{equation}

We now turn to the solution of the differential equation (\ref{potentialz1}), the nonhomogeneous differential equation for $\psi_1$. In order to obtain a particular solution to it we use the method of variation of parameters to write
\begin{equation}
    \psi_1(z) = 2\pi \sqrt{\omega z}J_{-\frac{j}{2}}(\omega z) \int \sqrt{\omega z}J_{\frac{j}{2}}(\omega z)\frac{\xi(z)}{W}dz - 2\pi \sqrt{\omega z}J_{\frac{j}{2}}(\omega z) \int \sqrt{\omega z} J_{-\frac{j}{2}}(\omega z)\frac{\xi(z)}{W}dz, \label{psi10}
\end{equation}
where $W$ denotes the wronskian of the basis of solutions of (\ref{psi0e})
\begin{equation}
   W= \frac{d}{dz}\left(\sqrt{2\pi}J_{-\frac{j}{2}}(\omega z)\right)\sqrt{2\pi}J_{\frac{j}{2}} - \frac{d}{dz}\left(\sqrt{2\pi}J_{\frac{j}{2}}(\omega z)\right)\sqrt{2\pi}J_{-\frac{j}{2}} = -4\omega \sin\left(\frac{\pi j}{2}\right).
\end{equation}
The extra function $\xi$ depends on $\psi_0$, given by (\ref{psi0}), and its derivatives, given by
\bea
\frac{d\psi_0}{dz}(z) &=& \sqrt{\frac{\pi}{2}}\sqrt{\frac{\omega}{z}}\left(A_+ J_{\frac{j}{2}}(z \omega )+A_+ \omega  z J_{\frac{j}{2}-1}(z \omega )-A_+ \omega  z J_{\frac{j}{2}+1}(z \omega )\right) \nonumber \\
&+& \sqrt{\frac{\pi}{2}}\sqrt{\frac{\omega}{z}}\left(A_- J_{-\frac{j}{2}}(z \omega )+A_- \omega  z J_{-\frac{j}{2}-1}(z \omega )-A_- \omega  z J_{1-\frac{j}{2}}(z \omega )\right), \\
\frac{d^2 \psi_0 }{d z^2}(z) &=& \sqrt{\frac{\pi }{8}} \sqrt{\omega} z^{-\frac{3}{2}} \left(j^2-4 \omega ^2 z^2-1\right) \left(A_+ J_{\frac{j}{2}}(z \omega )+A_- J_{-\frac{j}{2}}(z \omega )\right).
\eea
From (\ref{xi}) we see that $\xi$ also depends on three other quantities, whose leading terms in an arbitrarily small neighborhood around the origin are given by
\bea
V_1(r) &\sim& \frac{1}{4} (d-4)(d(d-5)(2 d-7)-22) \frac{R_H^{3d-7}}{r^{3d-5}}, \label{14} \\
f(r) \left[\delta f(r) \right]' &\sim& -\frac{(d-4)(d-3)(d-1)}{2} \frac{R_H^{2d-4}}{r^{2d-3}}, \\
2 \delta f(r) &\sim& -(d-3)(d-4) \left(\frac{R_H}{r}\right)^{d-1}.
\eea
Expressed in terms of $z$ through (\ref{z0}), the same leading terms are given by
\bea
2 \delta f(r) &\sim& \Upsilon_1 \left(\frac{R_H}{z}\right)^{-\rho-2}, \\
f(r) \left[\delta f(r) \right]' &\sim& \Upsilon_2 z^{\rho+1} R_H^{-\rho-2}, \\
V_1(r) &\sim& \Upsilon_3 z^\rho R_H^{-\rho-2}, \label{v10}
\eea
with the definitions
\bea
\rho &\equiv& -2-\frac{d-1}{d-2}, \label{rho}\\
\Upsilon_1 &\equiv& (-1)^{\rho+2} (d-2)^{\rho+2} (d-4)(d-3), \label{u1}\\
\Upsilon_2 &\equiv& (-1)^{\rho+1} (d-2)^{\rho+1} \frac{(d-4)(d-3)(d-1)}{2},  \label{u2}\\
\Upsilon_3 &\equiv& \frac{1}{4} (-1)^\rho (d-2)^\rho (d-4) ((d-5) d (2 d-7)-22) \label{u3}.
\eea
Analogously to (\ref{xi}), we decompose the particular solution to the nonhomogeneous differential equation (\ref{potentialz1}) into a sum of three terms, each one corresponding to one term of $\xi$:
\bea
\psi_1(z) &=& \sum_{i=1}^3 \phi_i(z), \nonumber \\
\phi_i(z) &=& 2\pi \sqrt{\omega z}J_{-\frac{j}{2}}(\omega z) \int \sqrt{\omega z}J_{\frac{j}{2}}(\omega z)\frac{\xi_i(z)}{W}dz - 2\pi \sqrt{\omega z}J_{\frac{j}{2}}(\omega z) \int \sqrt{\omega z} J_{-\frac{j}{2}}(\omega z)\frac{\xi_i(z)}{W}dz. \label{fi}
\eea

In appendix \ref{appendix1} we obtain explicit expressions for the functions $\phi_1(z), \phi_2(z), \phi_3(z)$. Based on these results, we can write the asymptotic behavior of $\psi_1$ in the limit $|\omega z| \gg 1$ as
\begin{equation}
    \psi_1(z) \sim \left(\Lambda_I^+e^{-i\alpha_+}+\Lambda_I^-e^{-i\alpha_-}\right)e^{i\omega z} + \left(\Lambda_I^+e^{i\alpha_+}+\Lambda_I^-e^{i\alpha_-}\right)e^{-i\omega z}, \label{psias1}
\end{equation}
with the definition
\begin{equation}
    \Lambda_I^\pm(d,j,\omega,R_H) = \left(R_H \omega \right)^{\frac{d-1}{d-2}} \sum_{k=1}^3\Theta_k^\pm(d,j,1,1) \label{lambi}
\end{equation}
and the coefficients $\Theta_{1,2,3}^\pm (d,j,\omega,R_H)$ (also depending on the constants $A_+, A_-$ from (\ref{psi0})) given at the end of appendix \ref{appendix1}. These coefficients all share the same overall factor $\left(R_H \omega \right)^{\frac{d-1}{d-2}}$. We will discuss this issue in section \ref{disc}.

We see that the above asymptotic behavior (\ref{psias1}) is not compatible with the boundary condition (\ref{bcz1}). In order to fix this we use the fact that we can always add a solution $\psi^*$ of the homogeneous equation associated with the differential equation (\ref{potentialz1}) to the \emph{particular} solution (\ref{psi10}). We use this fact to redefine $\psi_1$ as
\begin{equation}
    \psi_1 \to \psi_1 + \psi^*. \label{psiredef}
\end{equation}

The general solution of such homogeneous equation $\psi^*$ is, analogously to (\ref{psi0}), given by
\begin{equation} \label{psist}
    \psi^*(z) = C_+ \sqrt{2\pi}\sqrt{\omega z} J_{\frac{j}{2}}(\omega z) + C_-\sqrt{2\pi}\sqrt{\omega z}J_{-\frac{j}{2}}(\omega z)
\end{equation}
for some $C_+,C_- \in \mathbb{C}$. The asymptotic behavior of $\psi^*$ is also analogous to the one of $\psi_0$:
\be \label{psistas}
\psi^*(z) \sim \left( C_+ e^{-i\alpha_+} + C_- e^{-i\alpha_-}\right) e^{i \omega z} + \left( C_+ e^{i\alpha_+} + C_- e^{i\alpha_-}\right) e^{-i \omega z}.
\ee
We require $C_+,C_-$ to be solutions of the following linear system of equations (with $c \in \mathbb{C}$):
\begin{equation}
    \begin{cases}
    C_+e^{i\alpha_+} + C_-e^{i\alpha_-} = c, \\
    C_+e^{-i\alpha_+}+ C_-e^{-i\alpha_-} = -\left(\Lambda_I^+e^{-i\alpha_+}+\Lambda_I^-e^{-i\alpha_-}\right).
    \end{cases}
    \label{20}
\end{equation}
The second equation in (\ref{20}) is simply the condition for the $e^{i \omega z}$ term in (\ref{psistas}) to cancel the $e^{i \omega z}$ term in (\ref{psias1}) with the redefinition (\ref{psiredef}). Similarly to this redefinition, we may decompose the coefficients $C_\pm$ in (\ref{psist}) as
$$C_\pm=A_\pm+B_\pm,$$ with the coefficients $A_\pm$ satisfying (\ref{a0}) and $B_\pm$ satisfying
\begin{equation}
    \begin{cases}
    B_+e^{i\alpha_+} + B_-e^{i\alpha_-} = 0, \\
    B_+e^{-i\alpha_+}+ B_-e^{-i\alpha_-} = -\left(\Lambda_I^+e^{-i\alpha_+}+\Lambda_I^-e^{-i\alpha_-}\right),
    \end{cases}  \label{b0}
\end{equation}
whose solution is given by
\begin{equation}
    \begin{cases}
    B_- = \frac{e^{i\alpha_+}\left(\Lambda_I^+e^{-i\alpha_+}+\Lambda_I^-e^{-i\alpha_-}\right)}{2i\sin\left(\alpha_--\alpha_+\right)},  \\
    B_+ = -\frac{e^{i\alpha_-}\left(\Lambda_I^+e^{-i\alpha_+}+\Lambda_I^-e^{-i\alpha_-}\right) }{2i\sin\left(\alpha_--\alpha_+\right)}.
    \end{cases} \label{b0s}
\end{equation}
One can immediately see that, by adding the systems of equations (\ref{a0}) and (\ref{b0}), one obtains (\ref{20}). This means one can actually decompose $\psi^*$ in (\ref{psist}) into two parts. The coefficients of the first part are $A_\pm$ satisfying (\ref{a0}); having in mind (\ref{psidef}), the first part can actually be absorbed by $\psi_0$. Therefore, without loss of generality one can define $\psi^*$ as in (\ref{psist}), but with coefficients $B_\pm$ satisfying (\ref{b0}) (i.e. just by the second part of the splitting).

We get finally $\psi_1(z) \sim \left(\Lambda_I^+e^{i\alpha_+}+\Lambda_I^-e^{i\alpha_-}\right)e^{-i\omega z}$ (for the function after the redefinition (\ref{psiredef})) and, from (\ref{psidef}), (\ref{psias0}), (\ref{a0}), we obtain asymptotically as $|\omega z| \gg 1$
\begin{equation}
\begin{split}
    \psi(z) \sim \left(A_+e^{i\alpha_+} +A_-e^{i\alpha_-}\right)e^{-i\omega z} + \lambda' \left(\Lambda_I^+e^{i\alpha_+} +\Lambda_I^-e^{i\alpha_-}\right)e^{-i\omega z} \\ =
    \left(A_+e^{i\alpha_+}+A_-e^{i\alpha_-}\right)\left(1 + \lambda' \left(\frac{\Lambda_I^+e^{i\alpha_+} +\Lambda_I^-e^{i\alpha_-}}{A_+e^{i\alpha_+} +A_-e^{i\alpha_-}}\right)\right)e^{-i\omega z}.
    \end{split} \label{psias}
\end{equation}

\subsection{Computation of the total monodromy in the big contour}
\label{big}
\noindent

We now proceed with the analysis of the behavior of $\psi$ around the big contour described in section \ref{bigc} and depicted in figure \ref{fig2}.

In the contour we consider, we start from region $D$ in the lower part of the complex \emph{r}-plane (close to $r= - i \infty$ for $d=5$ and in the fourth quadrant for $d>5$), where $\psi(z)$ is given by (\ref{psias}). We follow the Stokes line towards the origin. In a neighborhood of the origin, we make a $\frac{3\pi}{d-2}$ rotation around the origin; we present the detailed study of the behavior of $\psi = \psi_0 + \lambda' \psi_1$ under such rotation in appendix \ref{appendix2}. We then follow again a portion of the contour coincident with a Stokes line and all the way to the upper portion of the contour back to complex infinity (region $U$), where the asymptotic values of $\psi_0, \psi_1$ are given by (\ref{psi03pi}) and (\ref{psi13pi}) respectively. We notice that both asymptotic values contain terms proportional to $e^{i\omega z}$ and to $e^{-i\omega z}$. This is to be expected from the WKB approximation.

Now we want to see how $\psi$ behaves in the large arc shaped portion of the big contour. We notice that this portion of the contour no longer coincides
with a Stokes line of the WKB approximation of the master equation (\ref{potentialz0}). Because of the condition
$\Im(\omega z)\gg 0,$
the term proportional to $e^{i\omega z}$ gives only an exponentially small contribution to $\psi$ in this region of the complex $r-$plane; therefore, as we abandon $U$ and start following this portion of the big contour this term can be modified by the small correction terms arising from the WKB approximation that have been neglected in the plane wave approximation. This way, we cannot trust the coefficient multiplying $e^{i\omega z}$, as we cannot expect it to stay the same after closing the big contour. However, even not knowing the coefficient of this term, we know it is exponentially small in the large arc, and can be discarded there.
Thus near $D$, after closing the big contour we will then have (from (\ref{psi03pi}) and (\ref{psi13pi}), with coefficients $\Lambda_F^\pm(d,j,\omega,R_H)$ given by (\ref{fi23pi}), and discarding the $e^{i\omega z}$ term)

\begin{equation}
 \label{psias3pi}
 \psi (z) \sim \left(A_+e^{5i\alpha_+} + A_-e^{5i\alpha_-}\right)e^{-i\omega z} + \lambda'\left(\Lambda_F^+e^{5i\alpha_+} +\Lambda_F^-e^{5i\alpha_-}\right)e^{-i\omega z}.
\end{equation}
This way, we use the plane wave approximation to obtain the coefficient multiplying $e^{-i\omega z}$ in $\psi$, and we assume that the corresponding monodromy is equivalent to the monodromy of $\psi$, i.e. that the monodromy of $\psi$ is multiplicative. This assumption will be fully justified when we compute the monodromy of $\psi$ in the small contour, in section \ref{smc}.

Before computing the total monodromy of $\psi$, we must consider one further detail. As we mentioned in section \ref{coord}, as a function $z(r)$ has a branch point in each zero of $f_0(r)$: the real and the ``fictitious'' horizons (\ref{hors}). From the discussion in section \ref{topst}, the big contour does not enclose any of these ``fictitious'' horizons but, since it encloses the real horizon, a full loop around it is bound to cross a branch cut somewhere. Thus the expression (\ref{psias3pi}) above is written with respect to a variable $z$, defined on a branch, in the Riemann surface of $z$, different from the branch in which the variable $z$ used in (\ref{psias}) is defined. In order to relate these two variables we simply need to compute the monodromy of $z$ associated with one clockwise loop around the event horizon $r=R_H$.

Close to the horizon, from (\ref{tort2}) we write (ignoring an irrelevant integration constant)
\be
z=\int \frac{dr}{f_0(r)} \approx \frac{1}{f'_0(R_H)} \ln\left(1 -\frac{r}{R_H}\right) = \frac{R_H}{d-3} \ln\left(1 -\frac{r}{R_H}\right). \label{zhor}
\ee
From the parametrization $1 -\frac{r}{R_H} = \rho e^{i\theta}$ for $\rho \geq 0$ and $0 \leq \theta < 2\pi$, we can rewrite the expansion (\ref{zhor}) near $r=R_H$ as $z(r) \sim \frac{R_H}{d-3} \left(\ln\left(\rho\right) + i \theta\right).$ Following a full clockwise loop around $r=R_H$ is equivalent to letting $\theta$ run from $2\pi$ to $0$. Therefore, the monodromy of $z$, associated with a full clockwise loop around $r = R_H$, is
\begin{equation}
  m :=  -2\pi i \frac{R_H}{d-3}. \label{mm}
\end{equation}
Using this monodromy, we can relate the previously mentioned $z$ variables by redefining the one used in (\ref{psias3pi}) as $z \mapsto z +m.$ We can then consider the monodromy of $e^{-i\omega z}$ around the big contour and rewrite (\ref{psias3pi}) near $D$ as
\begin{equation}
    \psi(z) \sim \left(A_+e^{5i\alpha_+} + A_-e^{5i\alpha_-}\right)e^{-i\omega  m}e^{-i\omega z}\left[1 + \lambda' \left(\frac{\Lambda_F^+e^{5i\alpha_+} +\Lambda_F^-e^{5i\alpha_-}}{A_+e^{5i\alpha_+}+A_-e^{5i\alpha_-}}\right)\right]. \label{psias3pi2}
\end{equation}
Comparing (\ref{psias}) and (\ref{psias3pi2}), we can finally write the final monodromy of $\psi$ around the big contour as
\begin{equation}
    \mathcal{M}_1 = \left(\frac{A_+e^{5i\alpha_+} + A_-e^{5i\alpha_-}}{A_+e^{i\alpha_+} + A_-e^{i\alpha_-}}\right) e^{-i\omega m} \left(1 + \lambda' \delta \mathcal{M}_1\right)\label{dm1}
\end{equation}
where, to first order in $\lambda'$,
\begin{equation}
    \delta \mathcal{M}_1 = \frac{\Lambda_F^+e^{5i\alpha_+} +\Lambda_F^-e^{5i\alpha_-}}{A_+ e^{5i\alpha_+}+ A_-e^{5i\alpha_-}} - \frac{\Lambda_I^+e^{i\alpha_+} +\Lambda_I^-e^{i\alpha_-}}{A_+e^{i\alpha_+} + A_+e^{i\alpha_-}}.\label{dm}
\end{equation}
We recall that $\alpha_{\pm}$ are given by (\ref{alfa}); $A_\pm$ are given by (\ref{a0s}); $\Lambda_I^\pm(d,j,\omega,R_H)$ are given by (\ref{lambi}) with coefficients $\Theta_{1,2,3}^\pm(d,j,\omega,R_H)$ defined in appendix \ref{appendix1}; $\Lambda_F^\pm(d,j,\omega,R_H)$ are given by (\ref{fi23pi}) with coefficients $B_\pm$ given by (\ref{b0s}) and $\Xi_{1,2,3}^\pm(d,j,\omega,R_H)$ defined in appendix \ref{appendix2}.

One can check that, taking the limit $j\to 0$,
\begin{equation}
\frac{A_+e^{5i\alpha_+} + A_-e^{5i\alpha_-}}{A_+e^{i\alpha_+} + A_-e^{i\alpha_-}} = -3. \label{m3}
\end{equation}

Also in the limit $j\to 0$, and after a lot of algebraic manipulations, we get for (\ref{dm})
\bea
\delta\mathcal{M}_1&=&\left(\frac{R_H \omega}{d-2} \right)^{\frac{d-1}{d-2}} \mathrm{e}^{-\frac{2\pi i}{d-2}} \Pi_{\textsf{T}}(d), \label{126}\\
\Pi_{\textsf{T}}(d) &:=& \frac{2}{3} \frac{(d-4) \left(d(d-5)+2\right)}{d-1} \sqrt{\pi} \sin\left(\frac{\pi}{d-2}\right) \frac{\Gamma\left(\frac{1}{2(d-2)}\right) \Gamma\left(\frac{d-3}{2(d-2)}\right) }{\left[\Gamma\left(\frac{d-1}{2(d-2)}\right)\right]^2}. \nonumber
\eea
This way, we can rewrite (\ref{dm1}) as
\begin{equation}
    \mathcal{M}_1 = -3 e^{-i\omega m} \left(1 + \lambda' \delta \mathcal{M}_1\right). \label{dm2}
\end{equation}
\subsection{Computation of the monodromy in the small contour}
\label{smc}
\noindent

\begin{figure}[H]
\centering
\includegraphics[width=0.5\textwidth]{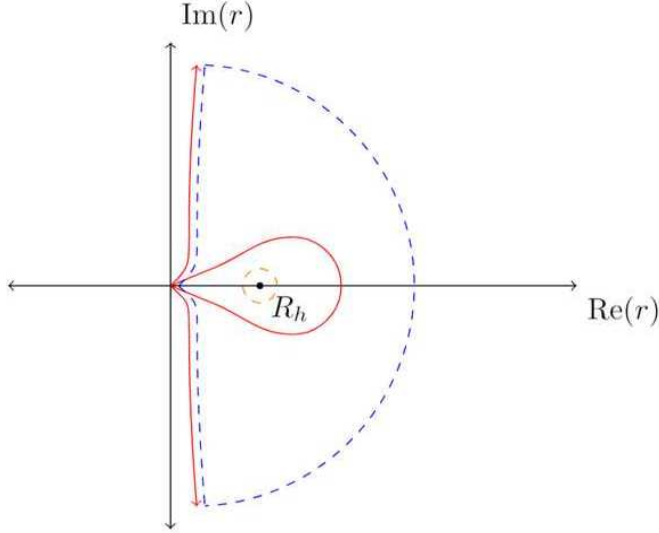}
\caption{Schematic depiction of the small and big contours as the orange and blue dashed lines respectively. The orange contour is to be interpreted as arbitrarily close to $R_H$. The Stokes lines are depicted by red curves. Naturally, not all Stokes lines are depicted.}
\label{fig4}
\end{figure}

Compared to the big one, the small contour is remarkably simple. Indeed, we build an arbitrarily small closed contour around the event horizon $R_H$. Such contour can be represented as the dashed orange contour in figure \ref{fig4}.

In this contour, we don't need to solve perturbatively the master equation (\ref{potential0}). Indeed, since the contour is arbitrarily small, we can simply solve (\ref{potential0}) expanded in a neighborhood of the event horizon. From $f(R_H)=0$ we see that $V_{\textsf{min}} [f(r)]$ given by (\ref{v0}), and more generally $V_{\textsf{T}} [f(r)]$ given by (\ref{vt}), vanish at the horizon. This way, (\ref{potential0}) is written in this region simply as
\begin{equation}
\frac{d^2 \psi}{d x^2} + \omega^2 \psi=0,
\end{equation}
with general solution given by $\psi(x) = C_+e^{i\omega x} + C_-e^{-i\omega x}$ for some constants $C_+,C_-$. Compatibility with the boundary condition (\ref{bcx2}) requires the restriction $C_- = 0$; therefore we may write
\begin{equation}
\psi(x) \sim C_+e^{i\omega x} \label{psis}
\end{equation}
in the small contour.

Close to the horizon, from (\ref{fr2}) we expand to first order in $\lambda$ $f(r) \approx f'_0(R_H) \left(1+ \frac{\lambda}{R_H^2} \delta f(R_H) \right) (r-R_H)$, from which we get (using (\ref{zhor}) and also ignoring an irrelevant integration constant)
\be
x(r)=\int \frac{dr}{f(r)} \approx \frac{1}{f'_0(R_H)} \left(1- \frac{\lambda}{R_H^2} \delta f(R_H) \right) \ln\left(1 -\frac{r}{R_H}\right) \approx \left(1- \frac{\lambda}{R_H^2} \delta f(R_H) \right) z(r)\label{xhor}
\ee
in the small contour. Thus, analogously to $z$, the variable $x$ has a branch point at the event horizon $R_H$. The monodromy $n$ of $x$, associated with a full clockwise loop around the small contour, is related to the equivalent monodromy $m$ of $z$, given by (\ref{mm}), as
\begin{equation}
    n:= \left(1- \frac{\lambda}{R_H^2} \delta f(R_H) \right) m = \left(1 + \frac{\lambda'}{2}(d-4)(d-1)\right)m.
    \label{mn}
\end{equation}
From (\ref{psis}), the monodromy of $\psi$ associated with a full clockwise loop around the small contour is multiplicative and given by
\begin{equation}
\mathcal{M}_2 := e^{i\omega n}.
\label{m2}
\end{equation}
The monodromy theorem tells us that this monodromy of $\psi$ is the same for every contour that is homotopic to the small contour, namely any larger contour around the horizon $R_H$, where $\psi$ is given by a combination of $e^{-i\omega z}$ and $e^{i\omega z}$. Thus the monodromy of (\ref{psis}), i.e. the monodromy of $e^{i\omega x}$, is indeed the monodromy of $\psi$. The multiplicativity of this monodromy is therefore a general property. The assumption of multiplicativity of the monodromy around the big contour is indeed justified, and so is its calculation in section \ref{big}.

\subsection{Equating monodromies}
\noindent

 Now, we want to relate the monodromies $\mathcal{M}_1$ and $\mathcal{M}_2$. To this end, we start by noticing that the big contour is homotopic to the small one. This is so, because one can continuously deform the big contour into the small one. Thus, using the monodromy theorem, we know that the monodromies of $\psi$, associated with the full clockwise loops around the big and the small contour, are the same.
 Hence, the equation
\begin{equation}
     \mathcal{M}_1 = \mathcal{M}_2
\end{equation}
must hold. From (\ref{dm2}) and (\ref{m2}) we can rewrite the equation above, to first order in $\lambda'$, as
\begin{equation}
-3 = e^{i\omega (m+n)}(1 - \lambda' \delta \mathcal{M}_1). \label{w3}
\end{equation}
Taking the logarithm on both sides of (\ref{w3}) we get, to first order in $\lambda'$,
\begin{equation}
    \ln(-3) - i \omega(m+n) + \lambda' \delta\mathcal{M}_1  = 0.
    \label{127}
\end{equation}

From (\ref{mm}) and (\ref{mn}), we can write
\begin{equation}
    m+n = -\left(2 + \lambda'\frac{(d-4)(d-1)}{2}\right)\frac{2\pi i}{d-3}R_H.
    \label{mmn}
\end{equation}
Using (\ref{temp}) we can rewrite $m+n$, to first order in $\lambda'$, as a function of the black hole temperature $T_{\mathcal{H}}$:
\begin{equation}
    m+n = -\frac{i}{T_{\mathcal{H}}}\left(1-\lambda'\frac{(d-1)(d-4)}{4}\right).
    \label{181}
\end{equation}
Similarly, from (\ref{temp}) we can also write, to first order in $\lambda'$,
\begin{equation}
    \lambda' \left(\frac{R_H\omega}{d-2}\right)^{\frac{d-1}{d-2}} = \lambda' \left[\frac{d-3}{d-2}\frac{\omega}{4\pi T_{\mathcal{H}}}\right]^{\frac{d-1}{d-2}}. \label{lrh}
\end{equation}
Taking (\ref{181}) for $m+n$, (\ref{126}) for $\delta \mathcal{M}_1$ and using (\ref{lrh}), we can rewrite (\ref{127}) as
\begin{equation}
    \ln(3) + (2k + 1)\pi i  = \frac{\omega}{T_\mathcal{H}}\left(1 - \lambda' \frac{(d-1)(d-4)}{4} - \lambda' \left[\frac{d-3}{d-2}\right]^{\frac{d-1}{d-2}} \left[\frac{\omega}{4\pi T_{\mathcal{H}}}\right]^{\frac{1}{d-2}} \frac{\Pi_{\textsf{T}}(d)}{4 \pi} \mathrm{e}^{-\frac{2}{d-2}\pi i} \right) \label{wtt1}
\end{equation}
for $k \in \mathbb{N}$. Here we take $k>0$ in order to get $\Im(\omega)>0$ in the Einstein limit $\lambda' \to 0$, and also considering condition (\ref{asy}). We obtain this way a transcendental equation that restricts the possible values of the quasinormal frequencies $\omega$ from the complex plane to an infinite but countable (for every value of the mode number $k$) subset. In the limit of Einstein gravity, we recover the results of \cite{Natario:2004jd}, as we should:
\be
\left. \frac{\omega}{T_\mathcal{H}}\right|_{\a=0} = \ln(3) + (2k + 1)\pi i, \, k \in \mathbb{N}. \label{qnme}
\ee
Since we are working perturbatively in $\lambda'$, we can simply consider this value of $\frac{\omega}{T_\mathcal{H}}$ for the $\lambda'$ correction in (\ref{wtt1}) and solve this equation for $\frac{\omega}{T_\mathcal{H}}$, obtaining
\begin{equation}
\frac{\omega}{T_\mathcal{H}} = \left(\ln(3) + (2k + 1)\pi i \right) \left(1 + \lambda' \frac{(d-1)(d-4)}{4} + \lambda' \left[\frac{d-3}{d-2}\right]^{\frac{d-1}{d-2}} \left[ \frac{\ln(3) + (2k + 1)\pi i}{4\pi}\right]^{\frac{1}{d-2}} \frac{\Pi_{\textsf{T}}(d)}{4 \pi} \mathrm{e}^{-\frac{2}{d-2}\pi i} \right). \label{wtt2}
\end{equation}
With this replacement, from (\ref{wtt2}) it is not easy to evaluate the real and imaginary parts of $\omega$. One should keep in mind that we have been working in the highly damped limit (\ref{asy}), which allows us to consider $\left.\frac{\omega}{T_\mathcal{H}}\right|_{\a=0} \approx (2k + 1)\pi i$ in the $\lambda'$ correction in (\ref{wtt2}). We also make the dependence on the temperature more explicit by writing $\lambda' = \lambda \frac{16 \pi^2}{(d-3)^2} T_\mathcal{H}^2$, according to (\ref{lprime}). This way we obtain for the $\a$-corrected asymptotic quasinormal spectra:
\begin{equation}
\frac{\omega}{T_\mathcal{H}} = \left(\ln(3) + (2k + 1)\pi i \right) \left[1 + \lambda \frac{16 \pi^2}{(d-3)^2} T_\mathcal{H}^2\left( \frac{(d-1)(d-4)}{4} + \left[\frac{d-3}{d-2}\right]^{\frac{d-1}{d-2}} \left[\frac{2k + 1}{4}\right]^{\frac{1}{d-2}} \frac{\Pi_{\textsf{T}}(d)}{4 \pi} \, \mathrm{e}^{-\frac{3 \pi i}{2(d-2)}}\right) \right], \label{wtt3}
\end{equation}
with $k \in \mathbb{N}$ and $\Pi_{\textsf{T}}(d)$ given by (\ref{126}).

Written in this form, for each value of $d$ one can evaluate the real and imaginary parts of the frequencies $\omega$. We notice that because of the $\mathrm{e}^{-\frac{3 \pi i}{2(d-2)}}$ term the $\a$ correction is complex, which means it will affect both the real and the imaginary parts of $\omega$. The real part of the asymptotic limit of $\frac{\omega}{T_\mathcal{H}}$, therefore, is no longer equal to the universal value $\ln(3)$: because of the $\a$-correction, it now also depends on the spacetime dimension $d$ and on the mode number $k$.

It is interesting to study the magnitudes of the different contributions to the $\lambda'$ correction in (\ref{wtt3}). For that purpose, we have
evaluated $\left[\frac{d-3}{d-2}\right]^{\frac{d-1}{d-2}} \frac{\Pi_{\textsf{T}}(d)}{4 \pi}$ numerically for the relevant values of $d$. This factor grows monotonically with $d$, varying from approximately 0.19 (corresponding to $d=5$) to approximately 13.5 (corresponding to $d=10$). Just for comparison, $\frac{(d-1)(d-4)}{4}$ varies between 1 and $13.5$ on the same range. For values of $k$ that are not very large (although verifying the condition $(2k + 1)\pi \gg \ln(3)$ defining the highly damped limit (\ref{asy})), the two terms in the $\lambda'$ correction in (\ref{wtt3}) are of comparable orders of magnitude. For these ``intermediate'' values of $k$ one must consider those two terms and, therefore, (\ref{wtt3}) cannot be simplified.

For fixed values of $\omega$, we notice our results may loose precision as $d$ increases. Indeed, as we discussed in appendix \ref{appendix1}, while computing the monodromy of $\psi$ around the big contour we made an approximation in the asymptotic expansion (\ref{expansion1}) by discarding terms of order at most $\left(\frac{\omega}{T_{\mathcal{H}}}\right)^{-\frac{1}{d-2}}.$ This is acceptable for the theoretical asymptotic limit, in which we assume that $|\omega| \to +\infty$. However, one must take care, when applying our results for very large, but not arbitrary so, values of $|\omega|$. Indeed, suppose one uses our transcendental equations to compute values of $\omega$ such that
\begin{equation}
 \Big\vert \frac{\omega}{T_{\mathcal{H}}} \Big\vert \sim 10^8.
\end{equation}
Then, those values were computed assuming that $10^{-\frac{8}{d-2}}$ is negligible. However, taking $d=5,7,9$ yields
\begin{equation}
     10^{-\frac{8}{3}} \approx 0.002, \hspace{20pt} 10^{-\frac{8}{5}} \approx 0.03, \hspace{20pt}
      10^{-\frac{8}{7}} \approx 0.07,
\end{equation}
respectively. As we can see, the approximation grows worst as we increase $d$. Thus, our results for progressively higher dimensions only apply to progressively higher values of $\Big\vert \frac{\omega}{T_{\mathcal{H}}} \Big\vert$.

\subsection{The asymptotic limit}
\label{disc}
\noindent

The decomposition of $\omega$ into its real and imaginary parts is simpler to obtain if one takes the asymptotic limit (i.e. the limit of large $k$) in (\ref{wtt3}).

One should keep in mind that we have been assuming in our calculations the highly damped limit (\ref{asy}). Assuming the result of Einstein gravity (\ref{qnme}), the highly damped limit (\ref{asy}) is simply equivalent to having $(2k + 1)\pi \gg \ln(3)$; a very large value of $k$ implies then the highly damped limit, but the converse does not need to be true. One may now ask the question: considering the $\a$ correction to $\omega$, does the large $k$ limit still imply the highly damped limit (\ref{asy})?

Throughout this work we have been taking a perturbative expansion in $\lambda$ (or $\a$). With such expansion, we have taken only terms of first order in $\lambda$, because that was the order considered in the lagrangian (\ref{eef}): terms of higher order are meaningless unless they had been included in the lagrangian since the beginning, changing the field equations and their solutions like (\ref{fr2}). These are only approximate and subject to even higher order corrections in $\lambda$, and the same is true for the perturbation potential $V_{\textsf{T}} [f(r)]$ given by (\ref{vt}). But once the order of perturbation is fixed and $V_{\textsf{T}} [f(r)]$ is given, the master equation (\ref{potential0}) is also fixed and can be solved exactly, like we did. The only reason we took the expansions (\ref{xint}) of the tortoise coordinate and (\ref{psidef}), (\ref{v1def}) of the perturbation function and potential was the fact that the $\lambda-$corrected master differential equation is nonhomogeneous, and the corresponding homogeneous equation is precisely the $\lambda=0$ master equation. The motivation for considering (\ref{psidef}), (\ref{v1def}) was not a perturbative expansion, but the (exact) method of variation of parameters to solve nonhomogeneous differential equations. This way, our solution to the master equation is exact to first order in $\lambda$, and so are the mathematical consequences that we can extract from it.

In general, a perturbative expansion means that a higher order term is expected to be negligible when compared to a lower order term, the reason being that the perturbative parameter is supposed to be small. That is certainly the case for our parameter $\lambda$. But that does not mean at all that $\lambda$ can be taken as arbitrarily small, and therefore it may happen that a higher order term of a perturbative expansion in $\lambda$ is not negligible if such term is made arbitrarily large.

But from equation (\ref{coef}) in appendix \ref{appendix1} and from (\ref{psias1}) and (\ref{lambi}) we see that $\lambda' \psi_1$ has an overall factor
$\left(R_H \omega \right)^{\frac{d-1}{d-2}}$ in the asymptotic limit when $|\omega z| \gg 1$. Such overall factor is also present in the $\lambda$ corrections $\delta\mathcal{M}_1$ to the monodromy $\mathcal{M}_1$ in (\ref{126}) and to the result for $\frac{\omega}{T_\mathcal{H}}$ in eq. (\ref{wtt1}) (in this case after having $R_H$ expressed in terms of $T_\mathcal{H}$ in (\ref{lrh})). One expects the perturbative parameter $\lambda'$ to be small, but if the absolute value $|\omega|$ is very large none of these $\lambda'$ corrections to $\psi$ and $\mathcal{M}_1$ is negligible, and nor is the one of $\omega$ itself. This is what happens in the asymptotic limit, with very large mode number $k$ and very large $\Im(\omega)$. One may argue that $\lambda'$ is a small parameter, and therefore the larger the order in $\lambda'$, the smaller the (relative) magnitude of the $\lambda'$ correction, even if their (absolute) magnitude is large. That argument may be true in our concrete discussion for large but finite $k$, but not in the asymptotic limit of very large $k$, say $k \to \infty$. For such limit the first order $\lambda'$ correction becomes too large, as one can see from (\ref{wtt2}) or from (\ref{wtt3}), and it cannot be neglected in any circumstance, contrarily to what one expects from a perturbative correction. Presumably the same is true for higher order $\lambda'$ corrections, which means in this limit the perturbative expansion breaks down. Therefore, strictly speaking, the asymptotic limit cannot be inferred from our result. The condition (\ref{asy}) we requested for the validity of our method should not be understood as ``arbitrarily large $\Im(\omega)$ (or $k$)'': $\Im(\omega)$ and $k$ can be very large, but not so large to break down the perturbative expansion. Other works computing asymptotic quasinormal modes with higher derivative corrections, such as \cite{Daghigh:2006xg}, require a similar interpretation for (\ref{asy}).

We suppose now that $k$ is large enough so that we can neglect the whole term multiplied by $\ln 3$ and also the term $\frac{(d-1)(d-4)}{4}$ in the $\lambda'$ correction in (\ref{wtt3}). If that is the case, we obtain then the simpler result that we were looking for in the beginning of this section:
\be
\frac{\omega}{T_\mathcal{H}} = \ln(3) + (2k + 1)\pi i
+ \lambda \left(\frac{4 \pi}{d-3}\right)^2 T_\mathcal{H}^2 \left[\frac{d-3}{d-2} \frac{(2k + 1)}{4}\right]^{\frac{d-1}{d-2}} \Pi_{\textsf{T}}(d) \, \mathrm{e}^{\frac{d-5}{2(d-2)}\pi i} \label{wtt4}
\ee
Similarly to (\ref{wtt3}), because of the $\mathrm{e}^{\frac{d-5}{2(d-2)}\pi i}$ term the $\a$ correction is complex, which means it will affect both the real and the imaginary parts of $\omega$. We have evaluated this term numerically for the relevant values of $d$. While $\cos\left(\frac{d-5}{2(d-2)}\pi\right)$ varies between 1 (corresponding to $d=5$) and approximately 0.56 (corresponding to $d=10$), $\sin\left(\frac{d-5}{2(d-2)}\pi\right)$ varies on the same range from 0 to approximately 0.83. For $d=5$, therefore, in this limit only the real part of $\omega$ gets an $\a$ correction. For other relevant values of $d$, (first order) $\a$ corrections to the real and the imaginary parts of $\omega$ are both positive and have similar orders of magnitude in this limit (for $d=8$ they are actually equal).

\section{Asymptotic quasinormal modes of test scalar fields}
\label{testas}
\noindent

A minimally coupled test scalar field propagating in the background of a black hole of the form (\ref{schwarz}) can be expanded as
\begin{equation}
\Phi(t,r,\theta)= e^{i\omega t} \sum_{\ell} \Phi_{\ell, \omega}(r) Y_{\ell}(\theta)\,,
\label{sphericalharmonics1}
\end{equation}
where $\omega$ is the wave frequency, $\ell$ is the angular quantum number associated with the polar angle $\theta$ and $Y_{\ell} (\theta)$ are the usual spherical harmonics defined over the $(d-2)$ unit sphere $\mathbb{S}^{d-2}$. Each component $\Phi_{\ell, \omega}(r)$ obeys a field equation like (\ref{potential0}), with precisely the same potential of the tensorial perturbations in Einstein gravity, given \cite{hep-th/0206084} by $V_{\textsf{min}} [f_0(r)]$ in (\ref{v0}), with $f_0(r)$ given by (\ref{fr0}). In the presence of $\a$ corrections this field equation remains the same (since the scalar field is minimally coupled); the only effect of the $\a$ corrections in it is indirect, through the metric. The potential corresponding to such equation is now given by $V_{\textsf{min}} [f(r)]$, with $f(r)$ corresponding to the $\a$-corrected metric (in our case given by (\ref{fr2})).

Naturally, if the field equation is the same, so is the spectrum of quasinormal modes. In Einstein gravity, therefore, test scalar fields and tensorial gravitational perturbations share the same spectra of quasinormal modes. For an $\a$-corrected metric, because of the difference in the potentials that is no longer true: the two spectra are indeed different. In \cite{Moura:2021eln} we verified this fact by computing the quasinormal modes in the eikonal limit for both cases.

In this section we address the calculation of the quasinormal modes corresponding to test scalar fields in the highly damped regime, like we previous did for tensorial gravitational perturbations, in the background of the $\a$-corrected black hole given by (\ref{fr2}). The calculation is totally analogous to the one described in section \ref{qnm}; we just have to consider the effects of changing the potential.

This change in the potential does not affect the monodromy $\mathcal{M}_2$ around the small contour, since close to the horizon the general potential $V_{\textsf{T}} [f(r)]$ vanishes, and so does the potential $V_{\textsf{min}} [f(r)]$ we are now considering. The same is true at infinity.

The result affected by the change in the potential, as one could expect, is the monodromy $\mathcal{M}_1$ around the big contour. The procedure we took in subsections \ref{pert} to \ref{big} to compute this monodromy, with a perturbative solution to the master equation, remains valid; it is actually totally analogous. The only change is in the expansion (\ref{v1def}) of the potential: $V_0(r)$ remains given by $V_{\textsf{min}} [f_0(r)]$, but because $V(r)$ is now different, so is $V_1(r)$.

Close to the origin, $V_1(r)$ is given by (\ref{v10}), but with $\Upsilon_3$ in (\ref{u3}) replaced by the new value
\begin{equation}
\Upsilon_3 \mapsto \Upsilon_3 = \frac{1}{4} (-1)^\rho (d-2)^{\rho+1} (d-4)(d-3)(2d-3). \label{191}
\end{equation}
This change in the value of $\Upsilon_3$ implies a change in the values of the coefficients $\Omega_{3}^\pm(d,j,\omega,R_H), \, \Theta_{3}^\pm(d,j,\omega,R_H)$ and $\Xi_{3}^\pm(d,j,\omega,R_H)$, and those changes will affect the values of $\Lambda_I^\pm(d,j,\omega,R_H)$, $\Lambda_F^\pm(d,j,\omega,R_H)$, and consequently of the $\lambda$ correction $\delta\mathcal{M}_1$ in the monodromy $\mathcal{M}_1$ around the big contour. This correction is now given by
\bea
\delta\mathcal{M}_1&=&\left(\frac{R_H \omega}{d-2} \right)^{\frac{d-1}{d-2}} \mathrm{e}^{-\frac{2\pi i}{d-2}} \Pi_{\textsf{S}}(d), \label{dms1}\\
\Pi_{\textsf{S}}(d) &:=& \frac{8}{3} \frac{(d-4) (d-3) (d-2)}{d-1} \frac{\pi^2}{2^{\frac{1}{d-2}}} \sin \left(\frac{\pi }{2(d-2)}\right) \frac{\Gamma \left(\frac{1}{d-2}\right) }{ \left[\Gamma \left(\frac{d-1}{2d-4}\right)\right]^4}. \nonumber
\eea

This expression has exactly the same form as the one previously obtained in (\ref{126}), just with $\Pi_{\textsf{T}}(d)$ replaced by $\Pi_{\textsf{S}}(d)$. All the results we previously obtained for the tensorial perturbations remain therefore valid for the test scalar field, just replacing $\Pi_{\textsf{T}}(d)$ by $\Pi_{\textsf{S}}(d)$. For the quasinormal frequencies we have then
\begin{equation}
\frac{\omega}{T_\mathcal{H}} = \left(\ln(3) + (2k + 1)\pi i \right) \left[1 + \lambda \frac{16 \pi^2}{(d-3)^2} T_\mathcal{H}^2\left( \frac{(d-1)(d-4)}{4} + \left[\frac{d-3}{d-2}\right]^{\frac{d-1}{d-2}} \left[\frac{2k + 1}{4}\right]^{\frac{1}{d-2}} \frac{\Pi_{\textsf{S}}(d)}{4 \pi} \, \mathrm{e}^{-\frac{3 \pi i}{2(d-2)}}\right) \right],
\end{equation}
with $k \in \mathbb{N}$ and $\Pi_{\textsf{S}}(d)$ given by (\ref{dms1}).

We have evaluated $\left[\frac{d-3}{d-2}\right]^{\frac{d-1}{d-2}} \frac{\Pi_{\textsf{S}}(d)}{4\pi}$ numerically for the relevant values of $d$. This factor grows monotonically with $d$, varying from approximately 0.58 (corresponding to $d=5$) to approximately 14.48 (corresponding to $d=10$). These values are of the same order of magnitude of those we obtained for the frequencies corresponding to tensorial perturbations (with $\Pi_{\textsf{T}}(d)$) in the discussion following (\ref{wtt3}). This means that all the conclusions we got in such discussion corresponding to tensorial perturbations (namely in section \ref{disc}) are also valid for test scalar fields. In particular the equivalent to the asymptotic limit (\ref{wtt4}) is also valid in this case in the same conditions it was obtained:
\be
\frac{\omega}{T_\mathcal{H}} = \ln(3) + (2k + 1)\pi i
+ \lambda \left(\frac{4 \pi}{d-3}\right)^2 T_\mathcal{H}^2 \left[\frac{d-3}{d-2} \frac{(2k + 1)}{4}\right]^{\frac{d-1}{d-2}} \Pi_{\textsf{S}}(d) \, \mathrm{e}^{\frac{d-5}{2(d-2)}\pi i}.
\ee

\section{Conclusions and future directions}
\noindent

In this work we have computed analytically the quasinormal frequencies in the highly damped limit corresponding to tensorial gravitational perturbations and scalar test fields for the simplest case of a $d-$dimensional spherically symmetric black hole solution with leading string $\a$ corrections given by the Callan-Myers-Perry black hole (\ref{fr2}). In both cases, we have obtained conditions that restrict the quasinormal frequencies $\omega$ to an infinite but countable set.

$\a$ corrections are expressed through a transcendental (in $\omega$) term in those conditions, that we may solve perturbatively in $\omega$, obtaining a closed form of the final result. In the Einstein limit, both the real and the imaginary parts of the quasinormal frequencies are independent of the spacetime dimension $d$, but the $\a$ corrections we obtained depend strongly on $d$. We checked that the magnitude of these $\a$ corrections increases with $d$.

The $\a$ corrections we obtained grow with the mode number $k$, which means the allowed values of $k$ should be large (in order to verify the highly damped limit we have assumed), but cannot be taken arbitrarily large, or the perturbative expansion will break down. Taking arbitrarily large values of $k$ does not have an immediate practical relevance; we considered that asymptotic limit only as a partial analysis. Assuming the higher order in $\a$ corrections also to grow with the mode number $k$, a complete analysis of the asymptotic limit should include the complete contributions from all such orders.

Since the value of $k$ is large but not arbitrarily large, corrections in powers of $1/k$ to the highly damped quasinormal frequencies could also be included. Corrections of that type have been studied in \cite{Musiri:2003bv, Cardoso:2003vt}, and those results should be considered together with the ones we obtain in this article.

In the highly damped limit, the $\a$ correction terms we obtained are different for quasinormal frequencies of tensorial gravitational perturbations and quasinormal frequencies of scalar test fields. Therefore, under the stringy correction considered, the highly damped limits of the associated quasinormal spectra are different. This situation is similar to the eikonal limit, where we also found different results for quasinormal frequencies of tensorial gravitational perturbations and scalar test fields \cite{Moura:2021eln}. However, in Einstein gravity all quasinormal frequencies (for tensorial, vectorial and scalar gravitational perturbations and for scalar test fields as well) are equal in the eikonal and in the highly damped limits. Stringy $\a$ corrections, therefore, allow us to distinguish between the quasinormal spectra of tensorial gravitational perturbations and scalar test fields in the eikonal and highly damped limits. It would be very interesting to check if vectorial and scalar gravitational perturbations also exhibit a similar behaviour in the stringy solution we considered. That is, should the quasinormal spectra associated with these perturbations be affected under the stringy correction as well and, if so, should they differ from the spectra associated with the remaining perturbations in the eikonal and highly damped limits?

In order to answer these questions, one must study vectorial and scalar gravitational perturbations in the Callan-Myers-Perry black hole and then obtain their quasinormal spectra in the highly damped and eikonal limits. These are topics for future works.

\paragraph{Acknowledgements}
\noindent
This work has been supported by Funda\c c\~ao para a Ci\^encia e a Tecnologia under contracts IT (UIDB/50008/2020 and UIDP/50008/2020) and project CERN/FIS-PAR/0023/2019.

\appendix

\section{The functions $\phi_1, \phi_2, \phi_3$}
\label{appendix1}
\noindent

We start by writing down explicit expressions for the functions $\phi_1(z), \phi_2(z), \phi_3(z)$ defined in (\ref{fi}), with $\rho$ defined in (\ref{rho}). $\phi_1(z)$ is given by
\bea
    \phi_1(z) &=& \left(\Omega_1^- A_-\right) \left(\sqrt{2\pi}\sqrt{\omega z}J_{-\frac{j}{2}}(\omega z)\right) \int \left(\sqrt{\omega z}J_{\frac{j}{2}}(\omega z)\right)\left(\sqrt{\omega z}J_{-\frac{j}{2}}(\omega z)\right)\left(j^2-4\omega^2z^2-1\right)z^{\rho}dz \nonumber \\ &+&
    \left(\Omega_1^+ A_-\right) \left(\sqrt{2\pi}\sqrt{\omega z}J_{\frac{j}{2}}(\omega z)\right) \int \left(\sqrt{\omega z}J_{-\frac{j}{2}}(\omega z)\right)\left(\sqrt{\omega z}J_{-\frac{j}{2}}(\omega z)\right)\left(j^2-4\omega^2z^2-1\right)z^{\rho}dz \nonumber \\&+&
    \left(\Omega_1^- A_+\right) \left(\sqrt{2\pi}\sqrt{\omega z}J_{-\frac{j}{2}}(\omega z)\right) \int \left(\sqrt{\omega z}J_{\frac{j}{2}}(\omega z)\right)\left(\sqrt{\omega z}J_{\frac{j}{2}}(\omega z)\right)\left(j^2-4\omega^2z^2-1\right)z^{\rho}dz \nonumber \\ &+&
    \left(\Omega_1^+ A_+\right) \left(\sqrt{2\pi}\sqrt{\omega z}J_{\frac{j}{2}}(\omega z)\right) \int \left(\sqrt{\omega z}J_{-\frac{j}{2}}(\omega z)\right)\left(\sqrt{\omega z}J_{\frac{j}{2}}(\omega z)\right)\left(j^2-4\omega^2z^2-1\right)z^{\rho}dz, \label{fi3}
\eea
with $\Upsilon_1$ defined in (\ref{u1}) and
\begin{equation}
    \Omega^\pm_1(d,j,\omega,R_H) = \pm \frac{\pi \Upsilon_1R_H^{\frac{d-1}{d-2}}}{8\omega}\csc\left(\frac{\pi j}{2}\right).
\end{equation}
$\phi_2(z)$ is given by
\bea
    \phi_2(z) &=& \left(\Omega_2^- A_-\right) \left(\sqrt{2\pi}\sqrt{\omega z}J_{-\frac{j}{2}}(\omega z)\right) \int \left(\sqrt{\omega z}J_{\frac{j}{2}}(\omega z)\right)\left(\sqrt{\omega z}J_{-\frac{j}{2}}(\omega z)\right)z^{\rho}dz \nonumber \\ &+&
    \left(\Omega_2^+ A_-\right) \left(\sqrt{2\pi}\sqrt{\omega z}J_{\frac{j}{2}}(\omega z)\right) \int \left(\sqrt{\omega z}J_{-\frac{j}{2}}(\omega z)\right)\left(\sqrt{\omega z}J_{-\frac{j}{2}}(\omega z)\right)z^{\rho}dz  \nonumber \\&+&
    \left(\Omega_2^- A_+\right) \left(\sqrt{2\pi}\sqrt{\omega z}J_{-\frac{j}{2}}(\omega z)\right) \int \left(\sqrt{\omega z}J_{\frac{j}{2}}(\omega z)\right)\left(\sqrt{\omega z}J_{\frac{j}{2}}(\omega z)\right)z^{\rho}dz \nonumber \\ &+&
    \left(\Omega_2^+ A_+\right) \left(\sqrt{2\pi}\sqrt{\omega z}J_{\frac{j}{2}}(\omega z)\right) \int \left(\sqrt{\omega z}J_{-\frac{j}{2}}(\omega z)\right)\left(\sqrt{\omega z}J_{\frac{j}{2}}(\omega z)\right)z^{\rho}dz \nonumber \\ &+&
    \left(\Omega_2^- A_-\right) \left(\sqrt{2\pi}\sqrt{\omega z}J_{-\frac{j}{2}}(\omega z)\right) \int \left(\sqrt{\omega z}J_{\frac{j}{2}}(\omega z)\right)\left(\sqrt{\omega z}J_{-\frac{j}{2}-1}(\omega z)\right)\omega z^{\rho+1}dz \nonumber \\ &+&
    \left(\Omega_2^+ A_-\right) \left(\sqrt{2\pi}\sqrt{\omega z}J_{\frac{j}{2}}(\omega z)\right) \int \left(\sqrt{\omega z}J_{-\frac{j}{2}}(\omega z)\right)\left(\sqrt{\omega z}J_{-\frac{j}{2}-1}(\omega z)\right)\omega z^{\rho+1}dz \nonumber \\&+&
    \left(\Omega_2^- A_+\right) \left(\sqrt{2\pi}\sqrt{\omega z}J_{-\frac{j}{2}}(\omega z)\right) \int \left(\sqrt{\omega z}J_{\frac{j}{2}}(\omega z)\right)\left(\sqrt{\omega z}J_{\frac{j}{2}-1}(\omega z)\right)\omega z^{\rho+1}dz \nonumber \\ &+&
    \left(\Omega_2^+ A_+\right) \left(\sqrt{2\pi}\sqrt{\omega z}J_{\frac{j}{2}}(\omega z)\right) \int \left(\sqrt{\omega z}J_{-\frac{j}{2}}(\omega z)\right)\left(\sqrt{\omega z}J_{\frac{j}{2}-1}(\omega z)\right)\omega z^{\rho+1}dz \nonumber \\ &-&
    \left(\Omega_2^- A_-\right) \left(\sqrt{2\pi}\sqrt{\omega z}J_{-\frac{j}{2}}(\omega z)\right) \int \left(\sqrt{\omega z}J_{\frac{j}{2}}(\omega z)\right)\left(\sqrt{\omega z}J_{-\frac{j}{2}+1}(\omega z)\right)\omega z^{\rho+1}dz \nonumber \\ &-&
    \left(\Omega_2^+ A_-\right) \left(\sqrt{2\pi}\sqrt{\omega z}J_{\frac{j}{2}}(\omega z)\right) \int \left(\sqrt{\omega z}J_{-\frac{j}{2}}(\omega z)\right)\left(\sqrt{\omega z}J_{-\frac{j}{2}+1}(\omega z)\right)\omega z^{\rho+1}dz \nonumber \\&-&
    \left(\Omega_2^- A_+\right) \left(\sqrt{2\pi}\sqrt{\omega z}J_{-\frac{j}{2}}(\omega z)\right) \int \left(\sqrt{\omega z}J_{\frac{j}{2}}(\omega z)\right)\left(\sqrt{\omega z}J_{\frac{j}{2}+1}(\omega z)\right)\omega z^{\rho+1}dz \nonumber \\ &-&
    \left(\Omega_2^+ A_+\right) \left(\sqrt{2\pi}\sqrt{\omega z}J_{\frac{j}{2}}(\omega z)\right) \int \left(\sqrt{\omega z}J_{-\frac{j}{2}}(\omega z)\right)\left(\sqrt{\omega z}J_{\frac{j}{2}+1}(\omega z)\right)\omega z^{\rho+1}dz, \label{fi2}
\eea
with $\Upsilon_2$ defined in (\ref{u2}) and
\begin{equation}
    \Omega^\pm_2(d,j,\omega,R_H) = \pm \frac{\pi \Upsilon_2R_H^{\frac{d-1}{d-2}}}{4\omega}\csc\left(\frac{\pi j}{2}\right).
\end{equation}
Finally $\phi_3(z)$ is given by
\bea
\phi_3(z) &=& \left(\Omega_3^- A_-\right) \left(\sqrt{2\pi}\sqrt{\omega z}J_{-\frac{j}{2}}(\omega z)\right) \int \left(\sqrt{\omega z}J_{\frac{j}{2}}(\omega z)\right)\left(\sqrt{\omega z}J_{-\frac{j}{2}}(\omega z)\right)z^{\rho}dz \nonumber \\ &+&
    \left(\Omega_3^+ A_-\right) \left(\sqrt{2\pi}\sqrt{\omega z}J_{\frac{j}{2}}(\omega z)\right) \int \left(\sqrt{\omega z}J_{-\frac{j}{2}}(\omega z)\right)\left(\sqrt{\omega z}J_{-\frac{j}{2}}(\omega z)\right)z^{\rho}dz  \nonumber \\&+&
    \left(\Omega_3^- A_+\right) \left(\sqrt{2\pi}\sqrt{\omega z}J_{-\frac{j}{2}}(\omega z)\right) \int \left(\sqrt{\omega z}J_{\frac{j}{2}}(\omega z)\right)\left(\sqrt{\omega z}J_{\frac{j}{2}}(\omega z)\right)z^{\rho}dz \nonumber \\ &+&
    \left(\Omega_3^+ A_+\right) \left(\sqrt{2\pi}\sqrt{\omega z}J_{\frac{j}{2}}(\omega z)\right) \int \left(\sqrt{\omega z}J_{-\frac{j}{2}}(\omega z)\right)\left(\sqrt{\omega z}J_{\frac{j}{2}}(\omega z)\right)z^{\rho}dz, \label{fi1}
\eea
with $\Upsilon_3$ defined in (\ref{u3}) and
\begin{equation}
    \Omega_3^\pm(d,j,\omega,R_H) =\pm \frac{\pi\Upsilon_3 R_H^{\frac{d-1}{d-2}}}{2\omega} \csc\left(\frac{\pi j}{2}\right).
\end{equation}
In order to evaluate these functions, we need to study the following class of indefinite integrals:
\begin{equation}
\mathcal{P}_{m n k}\left(x\right) = \int x^k J_m(x)J_n(x)dx, \label{pkmn}
\end{equation}
for $m,n \in \mathbb{R}$ and $k < 0$. These integrals evaluate to
\begin{equation}
   \mathcal{P}_{m n k}\left(x\right) =  \frac{2^{-m -n } x^{k +m +n +1}\, _3F_4\left(\frac{m+n+1}{2},\frac{m+n+2}{2},\frac{k+m+n+1}{2};m +1,\frac{k+m+n+3}{2},n +1,m +n +1;-x^2\right)}{\Gamma (m +1) \Gamma (n +1) (k +m +n +1)}, \label{hkmng}
\end{equation}
where
$$\,_3F_4\left(a,b,c;d,e,f,g;w\right) = \sum_{n = 0}^{+\infty}\frac{\Gamma(a+n)\Gamma(b+n)\Gamma(c+n)\Gamma(d)\Gamma(e)\Gamma(f)\Gamma(g)}{\Gamma(a)\Gamma(b)\Gamma(c)\Gamma(d+n)\Gamma(e+n)\Gamma(f+n)\Gamma(g+n)} \frac{w^n}{n!}$$
is a generalized hypergeometric function. This expression has the following asymptotic behavior for $|x|\gg 1$, considering that $k < 0$:
\begin{equation}
    \mathcal{P}_{ m nk}\left(x\right) \sim \mathcal{H}(m,n,k) +\mathcal{O}(x^k),
    \label{expansion1}
\end{equation}
where
\begin{equation}
    \mathcal{H}(m,n,k):= \frac{\Gamma \left(\frac{1}{2}-\frac{k }{2}\right) \Gamma \left(-\frac{k }{2}\right) \Gamma \left(\frac{k }{2}+\frac{m }{2}+\frac{n }{2}+\frac{1}{2}\right)}{2 \sqrt{\pi } \Gamma \left(-\frac{k }{2}+\frac{m }{2}-\frac{n }{2}+\frac{1}{2}\right) \Gamma \left(-\frac{k }{2}+\frac{n }{2}-\frac{m }{2}+\frac{1}{2}\right) \Gamma \left(-\frac{k}{2}+\frac{m }{2}+\frac{n }{2}+\frac{1}{2}\right)}. \label{hkmn}
\end{equation}
Finally, as a first approximation we can use the above asymptotic expansion together with $(\ref{alfa})$ to obtain the asymptotic behavior of $\phi_1(z), \phi_2(z), \phi_3(z)$ for $|\omega z| \gg 1$:
\begin{equation}
    \phi_p(z) \sim \left( \Theta_p^+e^{i\alpha_+}+\Theta_p^-e^{i\alpha_-}\right)e^{-i\omega z} +\left( \Theta_p^+e^{-i\alpha_+}+\Theta_p^-e^{-i\alpha_-}\right)e^{i\omega z}, \, p=1, 2,3. \label{fikas}
\end{equation}
The coefficients $\Theta_{1,2,3}^\pm$ are given in terms of $\Omega_{1,2,3}^\pm$ and of the asymptotic coefficients $\mathcal{H}$ defined in (\ref{hkmn}) by
\begin{equation}
\begin{split}
    \Theta_1^+(d,j,\omega,R_H) := \Omega^+_1 \omega^{-\rho-1}\left(A_-\mathcal{H}\left(-\frac{j}{2},-\frac{j}{2},\rho +1\right) + A_+ \mathcal{H}\left(-\frac{j}{2},\frac{j}{2},\rho +1\right)\right)\left(j^2-1\right)\\  -4\Omega^+_1 \omega^{-\rho-1}\left(A_-\mathcal{H}\left(-\frac{j}{2},-\frac{j}{2},\rho +3\right) + A_+ \mathcal{H}\left(-\frac{j}{2},\frac{j}{2},\rho +3\right)\right),
    \end{split}
\end{equation}

\begin{equation}
\begin{split}
    \Theta_1^-(d,j,\omega,R_H) := \Omega^-_1 \omega^{-\rho-1}\left(A_-\mathcal{H}\left(\frac{j}{2},-\frac{j}{2},\rho +1\right) + A_+ \mathcal{H}\left(\frac{j}{2},\frac{j}{2},\rho +1\right)\right)\left(j^2-1\right) \\  -4\Omega^-_1 \omega^{-\rho-1}\left(A_-\mathcal{H}\left(\frac{j}{2},-\frac{j}{2},\rho +3\right) + A_+ \mathcal{H}\left(\frac{j}{2},\frac{j}{2},\rho +3\right)\right),
    \end{split}
\end{equation}

\begin{equation}
\begin{split}
   \Theta^+_2(d,j,\omega,R_H) := \Omega^+_2 \omega^{-\rho-1} \left(A_-\mathcal{H}\left(-\frac{j}{2},-\frac{j}{2},\rho+1\right) + A_+\mathcal{H}\left(-\frac{j}{2},\frac{j}{2},\rho+1\right)\right) \\ + \Omega^+_2\omega^{-\rho-1}\left(A_-\mathcal{H}\left(-\frac{j}{2},-\frac{j}{2}-1,\rho+2\right)+A_+\mathcal{H}\left(-\frac{j}{2},\frac{j}{2}-1,\rho+2\right)\right) \\
   -\Omega_2^+\omega^{-\rho-1}\left(A_-\mathcal{H}\left(-\frac{j}{2},1-\frac{j}{2},\rho+2\right) + A^+\mathcal{H}\left(-\frac{j}{2},\frac{j}{2}+1,\rho+2\right)\right),
   \end{split}
\end{equation}

\begin{equation}
\begin{split}
   \Theta^-_2(d,j,\omega,R_H) := \Omega^-_2 \omega^{-\rho-1} \left(A_-\mathcal{H}\left(\frac{j}{2},-\frac{j}{2},\rho+1\right) + A_+\mathcal{H}\left(\frac{j}{2},\frac{j}{2},\rho+1\right)\right) \\ + \Omega^-_2\omega^{-\rho-1}\left(A_-\mathcal{H}\left(\frac{j}{2},-\frac{j}{2}-1,\rho+2\right)+A_+\mathcal{H}\left(\frac{j}{2},\frac{j}{2}-1,\rho+2\right)\right) \\
   -\Omega_2^-\omega^{-\rho-1}\left(A_-\mathcal{H}\left(\frac{j}{2},1-\frac{j}{2},\rho+2\right) + A^+\mathcal{H}\left(\frac{j}{2},\frac{j}{2}+1,\rho+2\right)\right),
   \end{split}
\end{equation}

\begin{equation}
\Theta_3^+(d,j,\omega,R_H) := \Omega^+_3 \omega^{-\rho-1}\left(A_-\mathcal{H}\left(-\frac{j}{2},-\frac{j}{2},\rho +1\right) + A_+ \mathcal{H}\left(-\frac{j}{2},\frac{j}{2},\rho +1\right)\right),
\end{equation}

\begin{equation}
    \Theta_3^-(d,j,\omega,R_H) := \Omega^-_3 \omega^{-\rho-1}\left(A_-\mathcal{H}\left(\frac{j}{2},-\frac{j}{2},\rho +1\right) + A_+ \mathcal{H}\left(\frac{j}{2},\frac{j}{2},\rho +1\right)\right).
\end{equation}

After looking at these coefficients, we notice that discarding terms of order $\mathcal{O}(x^k)$ in the asymptotic expansion (\ref{expansion1}) amounted to discard terms proportional to at most $\omega^{\rho + 3}=\omega^{-\frac{1}{d-2}}$.

These coefficients share the same overall factor $\left(R_H \omega \right)^{\frac{d-1}{d-2}}$: they all obey the relation
\be
\Theta_{1,2,3}^\pm (d,j,\omega,R_H)= \left(R_H \omega \right)^{\frac{d-1}{d-2}} \Theta_{1,2,3}^\pm (d,j,1,1). \label{coef}
\ee
We notice that, since we took $k<0$ in the definition (\ref{hkmn}), $\Theta_1^\pm$ are defined only for $\rho<-3$.

\section{Monodromy of $\psi$ near the origin under a $3\pi$ rotation in $z$-plane}
\label{appendix2}
\noindent

Here we compute the monodromy of $\psi(z)$ under a $3\pi$ rotation around the origin of the complex $z$-plane, or analogously a $\frac{3\pi}{d-2}$ rotation around the origin in the complex $r$-plane. Under such rotation, the monodromy of the Bessel functions is
\begin{equation}
\sqrt{\pi e^{3\pi i}z}J_{\pm \frac{j}{2}}(\omega e^{3\pi i}z) = e^{6i\alpha_\pm }\sqrt{\pi z} J_{\pm \frac{j}{2}}(\omega z).
\label{19}
\end{equation}
It is easy to obtain the monodromy of $\psi_0$ from (\ref{psi0}) and (\ref{19}).
The asymptotic behavior of $\psi_0$, which was originally given by (\ref{psias0}), after the rotation is given by
\begin{equation}
\psi_0(z) \sim \left(A_+e^{5i\alpha_+}+A_-e^{5i\alpha_-}\right)e^{-i\omega z} + \left(A_+e^{7i\alpha_+}+A_-e^{7i\alpha_-}\right)e^{i\omega z}. \label{psi03pi}
\end{equation}

In order to obtain the monodromy of $\psi_1$ after the same rotation, we have to study the functions $\phi_1(z), \phi_2(z), \phi_3(z)$ given respectively by (\ref{fi1}), (\ref{fi2}), (\ref{fi3}) and the indefinite integral $\mathcal{P}_{mnk}(z)$ given by (\ref{pkmn}). From the result for this indefinite integral obtained at (\ref{hkmng}), we see that we can write
\begin{equation}
\mathcal{P}_{mnk}(z)= z^{k+m +n+1}\mathcal{A}(z).
\end{equation}
where $\mathcal{A}$ is an even and analytic function of $z$ near the origin. Therefore, after a $3\pi$ rotation around the origin of the complex $z$-plane, we can associate to $\mathcal{P}_{mnk}(z)$ the monodromy
\begin{equation}
    \mathcal{P}_{mnk}(z) \to e^{3i\pi\left(k+m +n+1\right)}z^{k+m +n+1}\mathcal{A}(z) = e^{3i\pi\left(m +n+k+1\right)}\mathcal{P}_{mnk}(z).
\end{equation}
We recall that asymptotically $\mathcal{P}_{mnk}(z)$ behaves as $\mathcal{H}_{mnk}$ given by (\ref{hkmn}), and $\phi_1, \phi_2, \phi_3$ behave as (\ref{fikas}).

After completing the $3\pi$ rotation around the origin of the complex $z$-plane, and taking into consideration the monodromy of the Bessel functions (\ref{19}),
$\phi_1(z), \phi_2(z), \phi_3(z)$ have the following asymptotic expansions for $|\omega z| \gg 1$:
\begin{equation}
    \phi_p(z) \sim \left( \Xi_p^+e^{5 i\alpha_+} +\Xi_p^-e^{5i\alpha_-}\right)e^{-i\omega z} +\left( \Xi_p^+e^{7i\alpha_+}+\Xi_p^-e^{7i\alpha_-}\right)e^{i\omega z}, \, p=1, 2,3. \label{fi13pi}
\end{equation}
where, using the coefficients $\Omega_{1,2,3}^\pm$ defined in appendix (\ref{appendix1}), we further define the coefficients
\bea
\Xi_1^+(d,j,\omega,R_H) &=& \Omega^+_1 \omega^{-\rho-1}\left(A_-e^{3\pi i (\rho+2-j)}\mathcal{H}\left(-\frac{j}{2},-\frac{j}{2},\rho +1\right) + A_+ e^{3\pi i (\rho+2)}\mathcal{H}\left(-\frac{j}{2},\frac{j}{2},\rho +1\right)\right) \left(j^2-1\right) \nonumber \\
&-& 4\Omega^+_1 \omega^{-\rho-1}\left(A_-e^{3\pi i(\rho+4-j)}\mathcal{H}\left(-\frac{j}{2},-\frac{j}{2},\rho +3\right) + A_+e^{3\pi i(\rho+4)} \mathcal{H}\left(-\frac{j}{2},\frac{j}{2},\rho +3\right)\right), \nonumber \\
\Xi_1^-(d,j,\omega,R_H) &=& \Omega^-_1 \omega^{-\rho-1}\left(A_-e^{3\pi i(\rho+2)}\mathcal{H}\left(\frac{j}{2},-\frac{j}{2},\rho +1\right) + A_+ e^{3\pi i(\rho+2+j)}\mathcal{H}\left(\frac{j}{2},\frac{j}{2},\rho +1\right)\right) \left(j^2-1\right) \nonumber \\
&-&4\Omega^-_1 \omega^{-\rho-1}\left(A_-e^{3\pi i(\rho+4)}\mathcal{H}\left(\frac{j}{2},-\frac{j}{2},\rho +3\right) + A_+e^{3\pi i(j+\rho+4)} \mathcal{H}\left(\frac{j}{2},\frac{j}{2},\rho +3\right)\right), \nonumber \\
\Xi^+_2(d,j,\omega,R_H) &=&  \Omega^+_2 \omega^{-\rho-1} \left(A_-e^{3\pi i(\rho + 2 -j)}\mathcal{H}\left(-\frac{j}{2},-\frac{j}{2},\rho+1\right) + A_+e^{3\pi i(\rho+2)}\mathcal{H}\left(-\frac{j}{2},\frac{j}{2},\rho+1\right)\right) \nonumber \\
&+& \Omega^+_2\omega^{-\rho-1}\left(A_-e^{3\pi i(\rho+2-j)}\mathcal{H}\left(-\frac{j}{2},-\frac{j}{2}-1,\rho+2\right)+A_+e^{3\pi i(\rho+2)}\mathcal{H}\left(-\frac{j}{2},\frac{j}{2}-1,\rho+2\right)\right) \nonumber \\
&-& \Omega_2^+\omega^{-\rho-1}\left(A_-e^{3\pi i(\rho+4-j)}\mathcal{H}\left(-\frac{j}{2},1-\frac{j}{2},\rho+2\right) + A_+e^{3\pi i(\rho+4)}\mathcal{H}\left(-\frac{j}{2},\frac{j}{2}+1,\rho+2\right)\right), \nonumber \\
\Xi^-_2(d,j,\omega,R_H) &=& \Omega^-_2 \omega^{-\rho-1} \left(A_-e^{3\pi i(\rho+2)}\mathcal{H}\left(\frac{j}{2},-\frac{j}{2},\rho+1\right) + A_+e^{3\pi i(j+\rho+2)}\mathcal{H}\left(\frac{j}{2},\frac{j}{2},\rho+1\right)\right) \nonumber \\
&+& \Omega^-_2\omega^{-\rho-1}\left(A_-e^{3\pi i (\rho+2)} \mathcal{H}\left(\frac{j}{2},-\frac{j}{2}-1,\rho+2\right) +A_+e^{3\pi i(j+\rho+2)}\mathcal{H}\left(\frac{j}{2},\frac{j}{2}-1,\rho+2\right)\right) \nonumber \\
&-&\Omega_2^-\omega^{-\rho-1}\left(A_-e^{3\pi i(\rho+4)}\mathcal{H}\left(\frac{j}{2},1-\frac{j}{2},\rho+2\right) + A_+e^{3\pi i(\rho + 4 +j)}\mathcal{H}\left(\frac{j}{2},\frac{j}{2}+1,\rho+2\right)\right), \nonumber \\
\Xi_3^+(d,j,\omega,R_H) &=& \Omega^+_3 \omega^{-\rho-1}\left(A_-e^{3i\pi(\rho-j+2)}\mathcal{H}\left(-\frac{j}{2},-\frac{j}{2},\rho +1\right) + A_+e^{3\pi i(\rho+2)} \mathcal{H}\left(-\frac{j}{2},\frac{j}{2},\rho +1\right)\right), \nonumber \\
\Xi_3^-(d,j,\omega,R_H) &=& \Omega^-_3 \omega^{-\rho-1}\left(A_-e^{3i\pi\left(\rho+2\right)}\mathcal{H}\left(\frac{j}{2},-\frac{j}{2},\rho +1\right) + A_+e^{3i\pi \left(\rho+2+j\right)} \mathcal{H}\left(\frac{j}{2},\frac{j}{2},\rho +1\right)\right).
\eea

We must consider the redefinition (\ref{psiredef}) and the contribution of $\psi^*$, whose asymptotic behavior after the $3\pi$ rotation is similar to the one of $\psi_0$, given by (\ref{psi03pi}), but with the coefficients $A_\pm$ replaced by $B_\pm$ given by (\ref{b0s}). This way we are led to the definition
\begin{equation}
    \Lambda^\pm_F(d,j,\omega,R_H) = \left(R_H \omega \right)^{\frac{d-1}{d-2}} \sum_{k=1}^3\Xi_k^\pm(d,j,1,1) + B_\pm \label{fi23pi}.
\end{equation}
Combining (\ref{fi13pi}), (\ref{fi23pi}) and the asymptotic behavior of $\psi^*$, we get for the asymptotic behavior of $\psi_1$ after a $3\pi$ rotation around the origin of the complex $z$-plane
\begin{equation}
    \psi_1 (z) \sim \left(\Lambda_F^+e^{5i\alpha_+}+\Lambda_F^-e^{5i\alpha_-}\right)e^{-i\omega z} +\left(\Lambda_F^+e^{7i\alpha_+}+\Lambda_F^-e^{7i\alpha_-}\right)e^{i\omega z}. \label{psi13pi}
\end{equation}


\begin{thebibliography}{10}
\bibitem{Berti:2009kk}
E.~Berti, V.~Cardoso and A.~O.~Starinets, \emph{Quasinormal modes of black holes and black branes},
Class. Quant. Grav. \textbf{26} (2009), 163001 \texttt{[arXiv:0905.2975 [gr-qc]]}.
\bibitem{Konoplya:2002zu}
R.~A.~Konoplya,
\emph{On quasinormal modes of small Schwarzschild-anti-de Sitter black hole},
Phys. Rev. \textbf{D66} (2002), 044009 \texttt{[arXiv:hep-th/0205142 [hep-th]]}.
\bibitem{Konoplya:2003ii}
R.~A.~Konoplya,
\emph{Quasinormal behavior of the d-dimensional Schwarzschild black hole and higher order WKB approach},
Phys. Rev. D \textbf{68} (2003), 024018 \texttt{[arXiv:gr-qc/0303052 [gr-qc]]}.
\bibitem{Konoplya:2011qq}
R.~A.~Konoplya and A.~Zhidenko, \emph{Quasinormal modes of black holes: From astrophysics to string theory},
Rev. Mod. Phys. \textbf{83} (2011), 793-836 \texttt{[arXiv:1102.4014 [gr-qc]]}.
\bibitem{Panotopoulos:2018hua}
G.~Panotopoulos,
\emph{Electromagnetic quasinormal modes of the nearly-extremal higher-dimensional Schwarzschild de Sitter black hole},
Mod. Phys. Lett. \textbf{A33} (2018) 23, 1850130 \texttt{[arXiv:1807.03278 [gr-qc]].}
\bibitem{Destounis:2020pjk}
K.~Destounis, R.~D.~B.~Fontana and F.~C.~Mena, \emph{Accelerating black holes: quasinormal modes and late-time tails},
Phys. Rev. \textbf{D102} (2020) 4, 044005 \texttt{[arXiv:2005.03028 [gr-qc]].}
\bibitem{Destounis:2018utr}
K.~Destounis, G.~Panotopoulos and \'A.~Rinc\'on, \emph{Stability under scalar perturbations and quasinormal modes of 4D Einstein Born Infeld dilaton spacetime: exact spectrum},
Eur. Phys. J. \textbf{C78} (2018) 2, 139 \texttt{[arXiv:1801.08955 [gr-qc]].}
\bibitem{Panotopoulos:2017hns}
G.~Panotopoulos and \'A.~Rinc\'on,
\emph{Quasinormal modes of black holes in Einstein-power-Maxwell theory},
Int. J. Mod. Phys. \textbf{D27} (2017) 03, 1850034 \texttt{[arXiv:1711.04146 [hep-th]]}.
\bibitem{Rincon:2018sgd}
\'A.~Rinc\'on and G.~Panotopoulos, \emph{Quasinormal modes of scale dependent black holes in (1+2)-dimensional Einstein-power-Maxwell theory},
Phys. Rev. \textbf{D97} (2018) 2, 024027 \texttt{[arXiv:1801.03248 [hep-th]]}.
\bibitem{Motl:2003cd}
L.~Motl and A.~Neitzke, \emph{Asymptotic black hole quasinormal frequencies}, Adv. Theor. Math. Phys. \textbf{7} (2003) 2, 307-330
\texttt{[hep-th/0301173]}.
\bibitem{Birmingham:2003rf}
D.~Birmingham, \emph{Asymptotic quasinormal frequencies of d-dimensional Schwarzschild black holes}, Phys. Lett. \textbf{B569} (2003), 199
\texttt{[arXiv:hep-th/0306004 [hep-th]]}.
\bibitem{Andersson:2003fh}
N.~Andersson and C.~J.~Howls, \emph{The Asymptotic quasinormal mode spectrum of nonrotating black holes}, Class. Quant. Grav. \textbf{21} (2004),
1623 \texttt{[gr-qc/0307020]}.
\bibitem{Musiri:2003ed}
S.~Musiri and G.~Siopsis, \emph{On Quasinormal modes of Kerr black holes}, Phys. Lett. \textbf{B579} (2004), 25 \texttt{[arXiv:hep-th/0309227 [hep-th]]}.
\bibitem{Cardoso:2004up}
V.~Cardoso, J.~Nat\'ario and R.~Schiappa, \emph{Asymptotic quasinormal frequencies for black holes in nonasymptotically flat space-times},
J. Math. Phys. \textbf{45} (2004), 4698 \texttt{[arXiv:hep-th/0403132 [hep-th]]}.
\bibitem{Musiri:2003rs}
S.~Musiri and G.~Siopsis, \emph{Asymptotic form of quasinormal modes of large AdS black holes}, Phys. Lett. \textbf{B576} (2003), 309
\texttt{[arXiv:hep-th/0308196 [hep-th]]}.
\bibitem{Natario:2004jd}
J.~Nat\'ario and R.~Schiappa, \emph{On the classification of asymptotic quasinormal frequencies for d-dimensional black holes and quantum gravity},
Adv.\ Theor.\ Math.\ Phys.\ {\bf 8} (2004) 6, 1001 \texttt{[hep-th/0411267]}.
\bibitem{Nollert:1993zz}
H.~P.~Nollert, \emph{Quasinormal modes of Schwarzschild black holes: The determination of quasinormal frequencies with very large imaginary parts},
Phys. Rev. \textbf{D47} (1993), 5253.
\bibitem{Cardoso:2003cj}
V.~Cardoso, R.~Konoplya and J.~P.~S.~Lemos, \emph{Quasinormal frequencies of Schwarzschild black holes in anti-de Sitter
space-times: A Complete study on the asymptotic behavior}, Phys. Rev. \textbf{D68} (2003), 044024 \texttt{[arXiv:gr-qc/0305037 [gr-qc]]}.
\bibitem{Berti:2003zu}
E.~Berti and K.~D.~Kokkotas, \emph{Asymptotic quasinormal modes of Reissner-Nordstrom and Kerr black holes}, Phys. Rev. \textbf{D68} (2003), 044027
\texttt{[arXiv:hep-th/0303029 [hep-th]].}
\bibitem{Berti:2003jh}
E.~Berti, V.~Cardoso, K.~D.~Kokkotas and H.~Onozawa, \emph{Highly damped quasinormal modes of Kerr black holes}, Phys. Rev. \textbf{D68} (2003),
124018 \texttt{[arXiv:hep-th/0307013 [hep-th]]}.
\bibitem{Blazquez-Salcedo:2016enn}
J.~L.~Bl\'azquez-Salcedo, C.~F.~B.~Macedo, V.~Cardoso, V.~Ferrari, L.~Gualtieri, F.~S.~Khoo, J.~Kunz and P.~Pani,
\emph{Perturbed black holes in Einstein-dilaton-Gauss-Bonnet gravity: Stability, ringdown, and gravitational-wave emission},
Phys. Rev. \textbf{D94} (2016) 10, 104024 \texttt{[arXiv:1609.01286 [gr-qc]]}.
\bibitem{Cano:2020cao}
P.~A.~Cano, K.~Fransen and T.~Hertog, \emph{Ringing of rotating black holes in higher-derivative gravity}, Phys. Rev. \textbf{D102} (2020) 4, 044047 \texttt{[arXiv:2005.03671 [gr-qc]]}.
\bibitem{Pierini:2021jxd}
L.~Pierini and L.~Gualtieri, \emph{Quasi-normal modes of rotating black holes in Einstein-dilaton Gauss-Bonnet gravity: the first order in rotation}
\texttt{[arXiv:2103.09870 [gr-qc]]}.
\bibitem{Moura:2021eln}
F.~Moura and J.~Rodrigues, \emph{Eikonal quasinormal modes and shadow of string-corrected $d$-dimensional black holes} \texttt{[arXiv:2103.09302 [hep-th]]}.
\bibitem{Daghigh:2006xg}
R.~G.~Daghigh, G.~Kunstatter and J.~Ziprick, \emph{The mystery of the asymptotic quasinormal modes of Gauss-Bonnet black holes},
Class. Quant. Grav. \textbf{24} (2007), 1981-1992 \texttt{[gr-qc/0611139]}.
\bibitem{ik03a}
A.~Ishibashi and H.~Kodama, \textit{A Master Equation for Gravitational Perturbations of Maximally Symmetric Black Holes in Higher Dimensions},
Prog.\ Theor.\ Phys.\ \textbf{110} (2003) 701 \texttt{[hep-th/0305147]}.
\bibitem{Moura:2006pz}
F.~Moura and R.~Schiappa, \textit{Higher-derivative corrected black holes: Perturbative stability and absorption cross-section in heterotic string theory}, Class.\ Quant.\ Grav.\ {\bf 24} (2007) 361 \texttt{[hep-th/0605001]}.
\bibitem{Moura:2012fq}
F.~Moura, \emph{Tensorial perturbations and stability of spherically symmetric $d$-dimensional black holes in string theory}, Phys.\ Rev.\ {\bf D87}
(2013), 044036 \texttt{[arXiv:1212.2904 [hep-th]]}.
\bibitem{cmp89}
C.~G.~Callan, R.~C.~Myers and M.~J.~Perry, \textit{Black Holes in String Theory}, Nucl.\ Phys.\ \textbf{B311} (1989) 673.
\bibitem{Moura:2009it}
F.~Moura, \emph{String-corrected dilatonic black holes in d dimensions}, Phys.\ Rev.\ {\bf D83} (2011), 044002 \texttt{[arXiv:0912.3051 [hep-th]]}.
\bibitem{hep-th/0206084}
V.~Cardoso and J.~P.~S.~Lemos, \textit{Black hole collision with a scalar particle in four-dimensional, five-dimensional and seven-dimensional anti-de Sitter space-times: Ringing and radiation}, Phys.\ Rev.\ {\bf D66} (2002) 064006 \texttt{[hep-th/0206084]}.
\bibitem{Musiri:2003bv}
S.~Musiri and G.~Siopsis, \emph{Perturbative calculation of quasinormal modes of Schwarzschild black holes}, Class. Quant. Grav. \textbf{20} (2003),
L285 \texttt{[arXiv:hep-th/0308168 [hep-th]]}.
\bibitem{Cardoso:2003vt}
V.~Cardoso, J.~P.~S.~Lemos and S.~Yoshida, \emph{Quasinormal modes of Schwarzschild black holes in four-dimensions and higher dimensions},
Phys. Rev. \textbf{D69} (2004), 044004 \texttt{[gr-qc/0309112]}.
\end{thebibliography}
\end{document}